\documentclass[twocol]{ametsoc}

\journal{jas}

\bibpunct{(}{)}{;}{a}{}{,}

\title{A New Interpretation of Vortex-Split Sudden Stratospheric Warmings in Terms of Equilibrium Statistical Mechanics}

\authors{Yuki Yasuda\correspondingauthor{Yuki Yasuda, Atmosphere and Ocean Research Institute, University of Tokyo, 5-1-5, Kashiwanoha, Kashiwa-shi, Chiba 277-8564, Japan}}

\affiliation{Atmosphere and Ocean Research Institute, University of Tokyo, Kashiwa, Japan}
\email{yuki.yasuda@17.alumni.u-tokyo.ac.jp}

\extraauthor{Freddy Bouchet and Antoine Venaille}
\extraaffil{Univ Lyon, Ens de Lyon, Univ Claude Bernard, CNRS, Laboratoire de Physique, F-69342 Lyon, France}

\abstract{
Vortex-split sudden stratospheric warmings (S-SSWs) are investigated by using the Japanese 55-year Reanalysis (JRA-55), a spherical barotropic quasi-geostrophic (QG) model, and equilibrium statistical mechanics. The QG model reproduces well the evolution of the composite potential vorticity (PV) field obtained from JRA-55 by considering a time-dependent effective topography given by the composite height field of the 550 K potential temperature surface. The zonal-wavenumber-2 component of the effective topography is the most essential feature required to observe the vortex splitting. The statistical-mechanics theory predicts a large-scale steady state as the most probable outcome of turbulent stirring, and such a state can be computed without solving the QG dynamics. The theory is applied to a disk domain, which is modeled on the north polar cap in the stratosphere. The equilibrium state is obtained by computing the maximum of an entropy functional. In the range of parameters relevant to the winter stratosphere, this state is anticyclonic. By contrast, cyclonic states are quasi-stationary states corresponding to saddle points of the entropy functional. The theoretical calculations are compared with the results of the quasi-static experiment in which the wavenumber-2 topographic amplitude is increased linearly and slowly with time. The results suggest that S-SSWs can be qualitatively interpreted as the transition from the cyclonic quasi-stationary state toward the anticyclonic equilibrium state. The polar vortex splits during the transition toward the equilibrium state. Without any forcing such as radiative cooling, the anticyclonic equilibrium state would be realized sufficiently after an S-SSW.
}

\begin{document}

\maketitle

\section{Introduction}

The Arctic polar vortex (hereafter, simply referred to as the polar vortex) in the winter stratosphere is one of the largest coherent vortices in the atmosphere \citep[e.g.,][]{Andrews:etal:1987, Haynes:2005, Waugh:Polvani:2010}. The polar vortex is a cyclonic flow maintained through the thermal wind relation by radiative cooling and is perturbed by planetary-scale Rossby waves generated by topography and land-sea contrast in the troposphere. Sufficiently strong wave forcings sometimes break down the polar vortex. This extreme event has a time scale of a few days and is accompanied by a reversal of zonal winds (from westerly to easterly) and a strong warming (sometimes over 50 K) in the polar stratosphere, called a major sudden stratospheric warming (SSW). The SSW is not only an interesting phenomenon but also practically important because it affects tropospheric eddies, i.e., weather systems \citep[e.g.,][]{Baldwin:Dunkerton:2001, Limpasuvan:etal:2004, Nakagawa:Yamazaki:2006, Charlton:Polvani:2007, Mitchell:etal:2013, Hitchcock:Simpson:2014, Kidston:etal:2015, Maycock:Hitchcock:2015}. SSWs are classified into vortex-displacement type (D-SSWs) and vortex-split type (S-SSWs) on the basis of the horizontal structure of the polar vortex \citep{Charlton:Polvani:2007}. In contrast with the D-SSW, during the S-SSW, the polar vortex has a nearly barotropic structure and splits into two daughter vortices almost simultaneously between the altitudes of about 20 and 40 km \citep{Matthewman:etal:2009}. Developing an understanding of the mechanism for SSWs remains an important theoretical subject. The aim of this study is to propose a minimal model consistent with a reanalysis dataset that allows for a new interpretation of S-SSWs with statistical mechanics arguments supported by numerical simulations.

Pioneering work on the mechanism for SSWs was made by \citet{Matsuno:1971}. He proposed that SSWs result from the amplification of wave forcings to the mean flows. Following \citet{Matsuno:1971}, SSWs have often been theoretically studied in the framework of wave-mean flow interactions. These studies may be roughly categorized into two groups. The first group employs a severely truncated dynamical system, called the Holton-Mass model \citep{Holton:Mass:1976}. The second group focuses on the resonance of Rossby waves and is hereafter referred to as the resonance theory  \citep[e.g.,][]{Tung:Lindzen:1979a, Tung:Lindzen:1979b}. Both types of studies rely on perturbation expansions in terms of a small parameter representing a normalized wave amplitude.

The Holton-Mass model describes interactions between a zonal mean flow, varying only in the vertical direction, and a single wave mode that also varies in the vertical direction and is generated by an effective bottom topography imitating tropospheric disturbances. As the topographic amplitude is increased, the model shows an abrupt transition from a quasi-steady state with a strong westerly wind (close to the radiative equilibrium) to a time-dependent state where the zonal wind vacillates between weak westerly and easterly \citep{Holton:Mass:1976}. \citet{Chao:1985} investigated the Holton-Mass model in terms of the catastrophe theory and argued that this transition (i.e., a catastrophe) corresponds to SSWs in the atmosphere. Similar abrupt transitions have been observed in a one-layer version of the Holton-Mass model \citep{Ruzmaikin:etal:2003, Birner:Williams:2008} and also in an improved Holton-Mass model in which the effective topography is specified through the Eliassen-Palm flux \citep{Sjoberg:Birner:2014}.

The second class of approaches, i.e., the resonance theories, may be further classified into linear and nonlinear theories. The linear theories have revealed that stationary Rossby waves can be resonant with topography in realistic velocity fields \citep{Tung:Lindzen:1979a, Tung:Lindzen:1979b, Esler:Scott:2005}. This means that an increase in the amplitude of an effective topography and also a resonance can lead to the wave amplification resulting in an SSW. 

When a fluid system is near the resonant state, the wave amplitude is large, and nonlinear effects can no longer be ignored. \citet{Plumb:1981a, Plumb:1981b} developed a weakly nonlinear theory based on a continuously stratified quasi-geostrophic (QG) model and showed that the nonlinear effect makes a positive feedback; that is, the nonlinear adjustment to the mean velocity occurs in a way that amplifies the wave. \citet{Matthewman:Esler:2011} developed Plumb's idea and investigated the onset of S-SSWs with a weakly nonlinear model. This model described the evolution of a barotropic vortex-Rossby wave generated by a wavenumber-2 effective topography. They showed that an abrupt transition to a large-amplitude solution occurred in the model when the topographic amplitude was over the threshold, which corresponded to the evolution of a Rossby wave in the onset of an S-SSW. One should also see \citet{Esler:Matthewman:2011} for the application of these theoretical methods to D-SSWs.

An abrupt change in a large-scale flow structure, such as SSWs, is reminiscent of a phase transition in statistical mechanics. The equilibrium statistical mechanics for geophysical flows has made great progress over the last two decades \citep[e.g.,][]{Salmon:1998, Majda:Wang:2006, Bouchet:Venaille:2012}. Statistical mechanics is a powerful tool suited for the analysis of a nonlinear system with many degrees of freedom (e.g., QG system). One can build phase diagrams for macroscopic properties of a flow (e.g., the direction of the flow or the number of vortices) by computing the most probable state among all possible configurations without solving the governing equations. The most probable state is given as the solution of a variational problem and is referred to as the equilibrium state. By studying such phase diagrams and the properties of the equilibrium states, it is possible to understand the effects of varying external parameters, such as total energy and topographic amplitude. The statistical mechanics can be used for a strongly nonlinear system because it does not rely on the assumption of small-amplitude perturbations (or wave-mean-flow decompositions). Conversely, the statistical mechanics is applied only to a freely evolving flow without forcing or dissipation and does not indicate anything about the time evolution of the system, such as the relaxation toward the equilibrium state. However, it may be a natural first step to examine the following two questions:
\begin{description}
\item[[Q1]] Is the mean state of the stratosphere accompanied by the polar vortex close to an equilibrium state?
\item[[Q2]] Is it possible to interpret S-SSWs (or D-SSWs) as a phase transition (i.e., a transition from an equilibrium state to another equilibrium state)?
\end{description}
This study tries to answer both questions.

The general theory of equilibrium statistical mechanics for geophysical flows is known as the Miller-Robert-Sommeria (MRS) theory \citep{Miller:1990, Robert:1991, Robert:Sommeria:1991}. The MRS theory takes into account all conserved quantities, which give the constraints for the variational problem, i.e., the total energy and any moment of potential vorticity (PV). For simplicity, we employ a subclass of the MRS theory as a useful guide to interpret the results of numerical experiments. This subclass of the equilibrium theory gives a variational problem in which an entropy is maximized under the constraints of two conserved quantities, the total energy and circulation (i.e., the first moment of PV) \citep{Chavanis:Sommeria:1996, Venaille:Bouchet:2009, Venaille:Bouchet:2011b, Naso:etal:2010}. Any solution of this sub-theory is a solution of the MRS theory, but the converse is not necessarily true \citep{Bouchet:2008}. This statistical-mechanics theory will be introduced in Section 4a.

Equilibrium statistical mechanics has already been applied to several geophysical fluid problems, including Jupiter's great red spots \citep{Bouchet:Sommeria:2002}, bottom trapped flows over oceanic bottom topographies \citep{Merryfield:1998, Venaille:2012}, mesoscale eddies and jets in the ocean \citep{Venaille:Bouchet:2011a}, an idealized jet in a two-layer QG model \citep{Esler:2008}, and hurricanes \citep{Prieto:etal:2001}. \citet{Prieto:Schubert:2001} investigated the final states of the polar vortex evolving from dynamically unstable initial states by using a spherical unforced barotropic model without effective topography.

This study proposes a new interpretation of S-SSWs based on equilibrium statistical mechanics. The S-SSW can be qualitatively understood as a transition from an entropy saddle point toward the entropy maximum. The term entropy refers to the number of possible (micro) states in a QG system. Independent of the statistical mechanics approach, we will also demonstrate that the salient features of the observed S-SSWs can be captured in the framework of the barotropic QG model, when the time-dependent effective topography is given by the height field of the 550 K potential temperature surface at about an altitude of 22 km. D-SSWs are not investigated here because a stratified fluid model is necessary to describe them \citep{Matthewman:etal:2009, Esler:Matthewman:2011}.

To the best of our knowledge, there has been no study on SSWs explicitly using thermodynamic potentials, such as entropy and Helmholtz free energy. A phase transition is defined as a transition from an equilibrium state (i.e., a global maximum/minimum of some thermodynamic potential) to another equilibrium state, or defined by a singularity of some thermodynamic potential. Although previous studies on SSWs may be implicitly influenced by equilibrium statistical mechanics, these studies do not rigorously show that SSWs can be interpreted as a phase transition \citep[e.g.,][]{Chao:1985, Yoden:1987a, Christiansen:2000, Monahan:etal:2003, Birner:Williams:2008, Matthewman:Esler:2011, Sjoberg:Birner:2014, Liu:Scott:2015}.

The present study is organized as follows. In Section 2, the spherical, barotropic QG model is constructed on the basis of the composite analysis of the Japanese 55-year Reanalysis (JRA-55). The validity of the QG model is verified through the direct comparison with the evolution of the composite PV field. In Section 3, to examine transitions among states of the polar vortex, a quasi-static experiment is performed, in which the amplitude of a wavenumber-2 effective topography is increased linearly and slowly with time. In Section 4, we apply the statistical-mechanics theory to a two-dimensional disk domain with wavenumber-2 topography. Although the disk domain is modeled on the polar stratosphere, Section 4 is somewhat independent and refers to few results in previous sections. In Section 5, using the theoretical calculations in Section 4, we interpret the results of the quasi-static experiment and then propose new interpretation of S-SSWs. Finally, the concluding remarks are given in Section 6.

\section{Validity of a spherical, barotropic quasi-geostrophic model}

In this section, we demonstrate that a spherical, barotropic quasi-geostrophic (QG) model reproduces well the evolution of the composite potential vorticity (PV) obtained from a reanalysis dataset (JRA-55).  Similar simple models have been used in previous studies on SSWs \citep[e.g.,][]{Polvani:Waugh:2004, Mirrokni:etal:2011, Matthewman:Esler:2011, Liu:Scott:2015, Scott:2016}. However, to the best of our knowledge, the present study gives the first confirmation of the relevance of the barotropic QG model through direct comparison with the reanalysis dataset.

\subsection{Construction of composite fields}

We analyze the Japanese 55-year Reanalysis [JRA-55, see \citet{Kobayashi:etal:2015} for details] and construct a composite S-SSW from 10 S-SSW events showing distinct splittings of the polar vortex. Following \citet{Seviour:etal:2013}, all S-SSWs are identified by applying the vortex-moment diagnostics to the geopotential at 10 hPa over the winters (December to March) of 1958/1959 to 2013/2014. In the vortex-moment diagnostics, the polar vortex is approximated to be an ellipse (called an equivalent ellipse) of uniform geopotential that is defined from the first and second moments of the geopotential field. An S-SSW requires the aspect ratio of the ellipse to remain larger than 2.3 for 7 consecutive days or more. The onset time is defined as the time when the threshold of 2.3 is first exceeded and is designated by $t=0$. To prevent counting the same S-SSW twice, once an S-SSW is identified, no event is defined within 30 days of the onset. 

Moreover, to extract S-SSWs associated with vortex splittings near the North Pole, we exclude events in which the ellipse centroid moves to the south of 66${}^\circ$ N once or more between $t=0$ and $7$ days. This procedure corresponds to excluding mixed S-SSWs, which have features common to S-SSWs and D-SSWs \citep{Mitchell:etal:2013}, because a D-SSW is defined as an event where the ellipse centroid remains equatorward of 66${}^\circ$ N for 7 consecutive days or more \citep{Seviour:etal:2013}. After this procedure, 18 S-SSWs are identified. About a half of them do not show distinct vortex splittings and are excluded.

To make a meaningful composite, an average of S-SSWs having similar characteristics needs to be determined. To this end, we further extract 10 S-SSWs in which the distance between the two ellipse centroids of the daughter vortices, after the vortex splitting, is 2000 km larger than the sum of the lengths of their major axes for at least 4 consecutive days. The following results are insensitive to the choice of the criterion values of 2000 km and 4 days. The same results were obtained when 1500 km and 3 days were used instead. Finally, as in \citet{Seviour:etal:2013}, each composite field is constructed by simply averaging over the 10 S-SSWs at each time $t$. The onset dates of the 10 S-SSWs are listed in Table \ref{tab01}.

\begin{table}[h]
	\caption{Onset dates of the 10 S-SSWs showing the distinct splittings of the polar vortex. For comparison, the corresponding onset dates in \citet{Seviour:etal:2013} are listed, where the symbol --- denotes that the 2012/13 winter is outside their analysis period.}\label{tab01}
\begin{center}
	\begin{tabular}{c|c}
		\topline
		This study & \citet{Seviour:etal:2013} \\
		\midline
		Jan. 03, 1968  & Dec. 29, 1967 \\
		Jan. 16, 1971  & Jan. 15, 1971 \\
		Feb. 04, 1973  & Feb. 04, 1973 \\
		Feb. 17, 1979  & Feb. 18, 1979 \\
		Dec. 25, 1984  & Dec. 25, 1984 \\
		Feb. 24, 1999  & Feb. 24, 1999 \\
		Mar. 15, 2001  & Mar. 15, 2001 \\
		Mar. 21, 2002  & Mar. 21, 2002 \\
		Jan. 18, 2009  & Jan. 18, 2009 \\
		Jan. 05, 2013  & --- \\
		\botline
	\end{tabular}
\end{center}
\end{table}

\begin{figure*}[!t]
	\centerline{\includegraphics[width=39pc,angle=0]{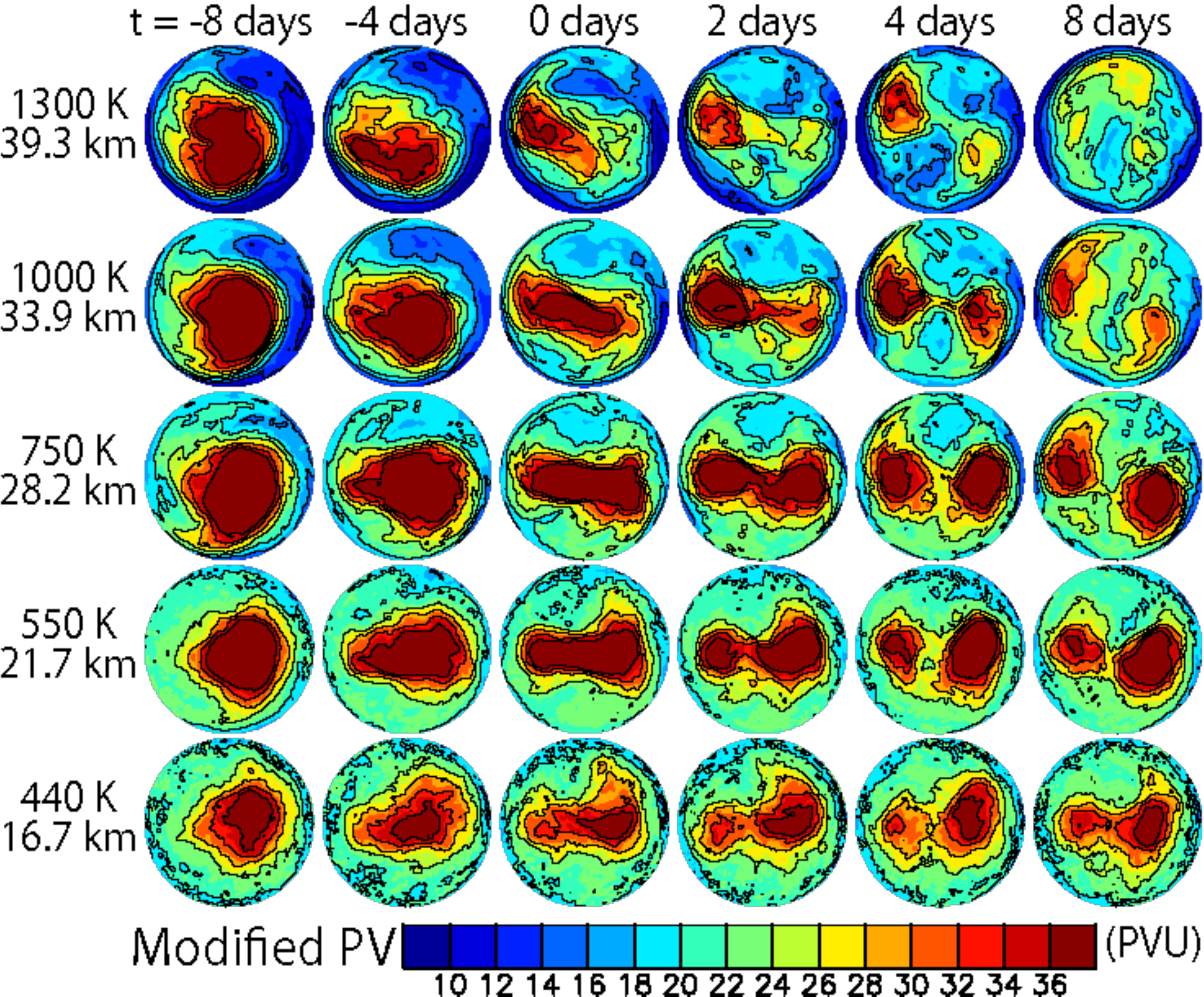}}
	\caption{
	Evolution of the composite modified PV over 30 to 90${}^\circ$ N at the respective potential temperature surfaces between 440 and 1300 K. The modified PV is defined as $\mathrm{EPV} \times (\theta / \theta_0)^{-9/2}$, where EPV is Ertel's PV, $\theta$ is the potential temperature, and $\theta_0 = 475$ K. The unit of PVU is defined as 1 PVU $=$ 10$^{-6}$ K m$^2$ s$^{-1}$ kg$^{-1}$. Each lateral label represents the time-mean height of the $\theta$ surface at 60${}^\circ$ N over the period of the composite S-SSW ($-10 \; \mathrm{days} \le t \le 10$ days).
	}\label{fig01}
\end{figure*}

The S-SSW is known to have a nearly barotropic structure \citep{Matthewman:etal:2009}. This barotropic structure is also observed in our composite S-SSW. Figure \ref{fig01} shows the snapshots of modified PV \citep{Lait:1994} at various times and heights. The modified PV is defined as $\mathrm{EPV} \times (\theta / \theta_0)^{-9/2}$, where EPV is Ertel's PV, $\theta$ is the potential temperature, and $\theta_0$ is its reference ($=$ 475 K). The modified PV is a conserved quantity along an air-parcel trajectory like Ertel's PV in the absence of friction and diabatic heating. Unlike Ertel's PV, however, the modified PV exhibits small vertical dependence, so that it is suitable to examine a vertical structure of the polar vortex. The polar vortex is approximately barotropic around $t=0$ and splits almost simultaneously between the altitudes of about 20 and 40 km. The vortex splitting is obscure at the 1300 K surface. After the polar vortex splits, the two daughter vortices are advected westward more rapidly while being weakened at high altitudes (1000 and 1300 K). These results are consistent with those of \citet{Matthewman:etal:2009}.

\subsection{Model description}

We describe the dynamics over the altitude range of about 20 to 40 km ($\theta =$ 550 to 1300 K) by using a spherical, barotropic QG model with an effective topography. The governing equations \citep[e.g.,][]{Vallis:2006} are
\begin{align}
	\partial_t q + \mathbf{v} \mathbf{\cdot \nabla} q  &= - \nu \Delta^{10} q, \text{ and} \\
	q(\mathbf{x}, t) &\equiv \Delta \psi(\mathbf{x}, t) + \underbrace{2 \Omega \sin \varphi}_{f(\mathbf{x})} \nonumber\\ &+ 2 \Omega \sin \varphi \; h(\mathbf{x}, t) / H - \underbrace{2 \Omega \sin \varphi_\mathrm{off}}_{f_\mathrm{off}},
\end{align}
where $q$ is the PV, $\psi$ is the stream function, $\mathbf{v}$ is the velocity given by $\mathbf{k} \times \mathbf{\nabla} \psi$ ($\mathbf{k}$ is a vertical unit vector), $\nu$ is a coefficient of hyperviscosity, $\Omega$ is the angular speed of the earth's rotation, $h$ is an effective bottom topography, and $H$ is an effective mean depth. Note that $h$ is considered to represent the effects of large-scale disturbances propagating from the troposphere, as in other theoretical models describing SSWs \citep[e.g., ][]{Matsuno:1971, Holton:Mass:1976}. A position $\mathbf{x}$ on the sphere is specified by a longitude $\lambda$ and a latitude $\varphi$. Without loss of generality, the offset of the PV, $f_\mathrm{off}$ ($\equiv 2 \Omega \sin \varphi_\mathrm{off}$), is introduced in (2), where $\varphi_\mathrm{off} = 45^\circ$ N. In the following, quantities except for time $t$ are basically nondimensionalized using the characteristic scales, such as one day, earth radius, and mean depth.

We numerically solve the governing equations (1) and (2) with the fourth-order Runge-Kutta method and the spectral method with the spherical harmonic expansion\footnote{The numerical model was constructed with a library (ISPACK) made by \citet{Ishioka:2013}.}. All experiments were conducted with the truncation wavenumber (in a triangular manner) T106 with a time step of 12 min. When calculating the nonlinear terms, we use the standard transform method with an alias-free grid of 320 (zonal) $\times$ 160 (meridional). Some experiments were also performed with T63, T85, and T126 to confirm the insensitivity to the truncation wavenumber. The value of $\nu$ was chosen such that the $e$-folding time for the modes with the highest total wavenumber was either 80 h, 8 h, or 2.4 h. We confirmed that all results are not highly sensitive to $\nu$. Unless otherwise stated, $H$ is set to the scale height of 6.14 km defined by $R_\mathrm{dry} T_0 / g$, where $R_\mathrm{dry}$ is the gas constant for dry air ($=$ 287 J K$^{-1}$ kg$^{-1}$), $g$ is the gravity acceleration ($=$ 9.81 m$^2$ s$^{-2}$), and $T_0$ ($=$ 210 K) is the climatological temperature for the midwinter over the polar cap north of 60$^\circ$ N in JRA-55.

\subsection{Reproduction of PV evolution associated with composite S-SSW}

To compare with the QG simulations, the composite PV is constructed from JRA-55 with the following three steps. (i) The effective topography $h$ is given at each time (every 6 h) by an undulation of the 550 K potential temperature surface (at about 22 km) and is denoted by $h_\mathrm{cmp}$. The undulation is defined as the height deviation of the 550 K surface from its mean height ($=$ 21.7 km), which is the zonal average at 60${}^\circ$ N over the period of the composite S-SSW ($-10 \; \mathrm{days} \le t \le 10$ days). (ii) The barotropic relative vorticity $\Delta \psi$ in (2) is obtained by vertically averaging the composite relative vorticity with a weight of density over $\theta = 550$ to $1300$ K (about 22 to 39 km). (iii) The composite PV $q_\mathrm{cmp}$ is determined by substituting the obtained $\Delta \psi$ and $h_\mathrm{cmp}$ into (2).

For the numerical experiments, $h_\mathrm{cmp}$ is linearly interpolated in time and then set to the QG model. The initial PV is given by the composite PV $q_\mathrm{cmp}$ at $t=-10$ days. Numerical integration is performed from $t=-10$ to $10$ days. All the following results are insensitive to the choice of the bottom potential temperature surface. Similar results were obtained when the 440 K surface (at about 17 km) was used instead of the 550 K surface.

\begin{figure*}[!t]
	\centerline{\includegraphics[width=39pc,angle=0]{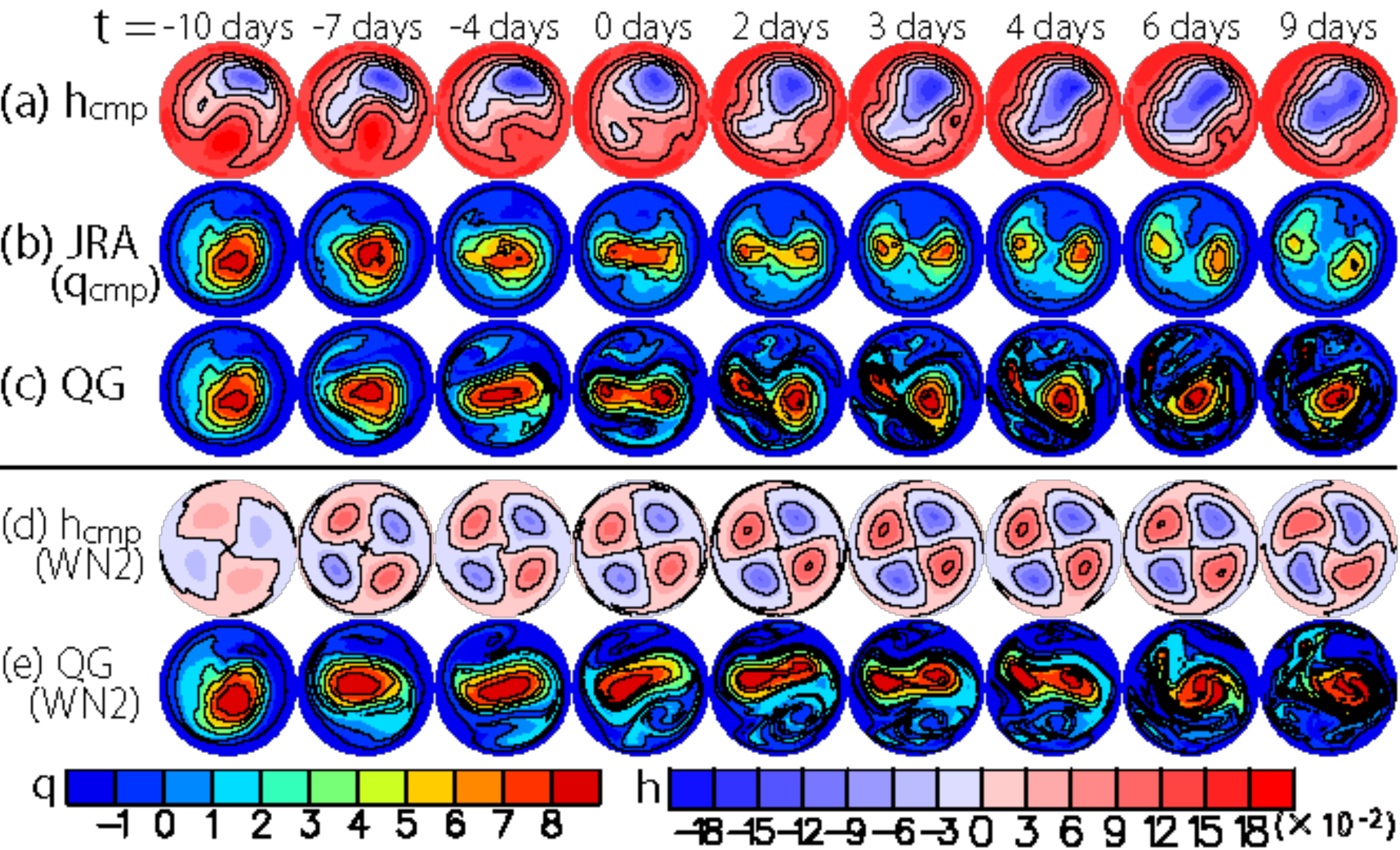}}
	\caption{
	Evolution over 0 to 90${}^\circ$ N of (a) the effective topography $h_\mathrm{cmp}$ from JRA-55, (b) the composite PV $q_\mathrm{cmp}$ from JRA-55, (c) the simulated PV $q$ by the QG model, (d) the zonal-wavenumber-2 component of $h_\mathrm{cmp}$, and (e) the simulated PV $q$ by the QG model with only the wavenumber-2 $h_\mathrm{cmp}$ shown in (d). The magnitude of $h_\mathrm{cmp}$ is nondimensionalized by $H = 6.14$ km in (a) and $0.7 \times 6.14$ km in (d).
	}\label{fig02}
\end{figure*}

Figures \ref{fig02}a, \ref{fig02}b, and \ref{fig02}c show the evolution of the effective topography $h_\mathrm{cmp}$, the composite PV $q_\mathrm{cmp}$, and the simulated PV $q$, respectively. The zonal-wavenumber-1 structure is observed in $h_\mathrm{cmp}$ at first, but gradually the wavenumber-2 structure becomes dominant. The PV evolution is well reproduced by the QG model until $t=4$ days, especially before and after the vortex splitting at about $t = 2$ days. At a later time (t = $6$ or $9$ days), the two daughter vortices remain in the composite PV field, but only one vortex remains in the simulated field and the other becomes quite small. This is likely due to the baroclinic structure that develops after the vortex splitting, as seen in Fig. \ref{fig01}.

Similar experiments were also conducted for the S-SSW in February 1979 \citep[e.g.,][]{Palmer:1981} instead of the composite S-SSW. This S-SSW event has been widely studied, because it occurred during a period of intensive observation of the atmosphere \citep[e.g.,][]{Andrews:etal:1987}. The PV evolution was also reproduced well by the QG model. Moreover, we found little influence of changing the initial time. A similar vortex splitting was simulated when the initial time was changed to $t=-20$ days (not shown).

We further examine the zonal-wavenumber component of $h_\mathrm{cmp}$ that is the most essential for reproducing the vortex splitting. Since the zonal wavenumber 0, 1, and 2 components are predominant in $h_\mathrm{cmp}$, only a single wavenumber component of $h_\mathrm{cmp}$ is given to the QG model, and the numerical integrations are conducted. In all cases, the deformation of the polar vortex is not clear, which suggests that the amplitude of each component of $h_\mathrm{cmp}$ is too small. As the use of a smaller mean depth $H$ results in a relatively larger amplitude of $h_\mathrm{cmp}$ in (2), all numerical simulations are performed again with $H = \tilde{c} \times 6.14$ km, where $\tilde{c}$ is set to $0.7$, $0.8$, or $0.9$. The vortex split occurs only in the three cases of the wavenumber-2 $h_\mathrm{cmp}$ with $\tilde{c}=0.7$, $0.8$, or $0.9$. Figures \ref{fig02}d and \ref{fig02}e show the evolution of the wavenumber-2 $h_\mathrm{cmp}$ and simulated $q$ with $\tilde{c}=0.7$, respectively. These results indicate that the zonal-wavenumber-2 component of $h_\mathrm{cmp}$ is the most essential feature of the effective topography that has to be retained in order to simulate the vortex splitting.

\section{Numerical experiments with a fixed spatial structure of effective topography}

In the numerical experiments in the previous section, the details of the transition (or bifurcation) of the polar vortex were not clear because the time scale of the effective topography was too short and the polar vortex evolved too rapidly. It may be useful for understanding S-SSWs to conduct an experiment with a slowly varying topography and to pursue changes in the structure of a flow field. In fact, several studies \citep[e.g.,][]{Matthewman:Esler:2011, Liu:Scott:2015} have argued that some transition is essential to interpret S-SSWs. To examine such a transition, we conduct a quasi-static experiment where the time scale of the effective topography $h$ is $O(10^4 \mathrm{days})$. Due to this quite slow variation of $h$, the flow field is nearly steady, and it is possible to reveal the states before and after a transition. The setup of the quasi-static experiment is also suitable for application of equilibrium statistical mechanics. The results will be interpreted in Section 5. 

\begin{figure*}[!t]
	\centerline{\includegraphics[width=39pc,angle=0]{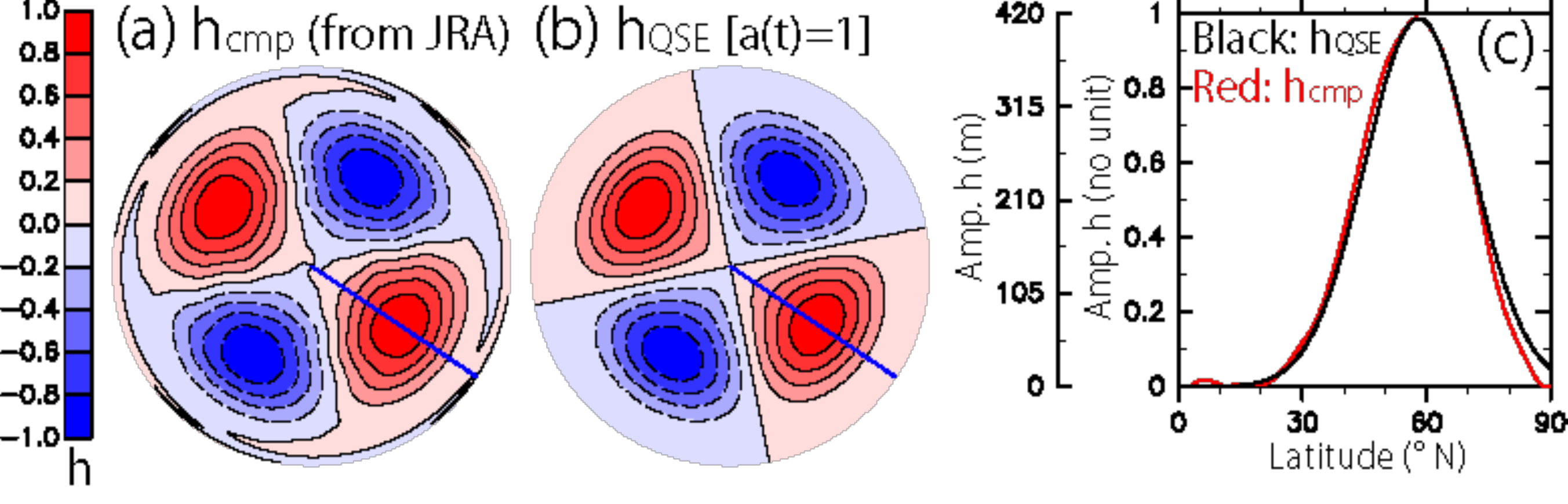}}
	\caption{
	Effective topographies over 0 to 90${}^\circ$ N: (a) the zonal-wavenumber-2 component of $h_\mathrm{cmp}$ at the onset time made from JRA-55 and (b) $h_\mathrm{QSE}$ with $a(t)=1$ defined in (3) and (4). The parameters $h_0$, $\lambda_0$, $\varphi_0$, and $\varDelta\varphi$ are determined to fit $h_\mathrm{QSE}$ into the wavenumber-2 $h_\mathrm{cmp}$: $h_0 = 420$ m, $\lambda_0=56.3^\circ$, $\varphi_0=58.1^\circ$ N, and $\varDelta\varphi = 13.0^\circ$. (c) Cross sections of (a) and (b) on the blue lines. In (a) through (c), the magnitude of the effective topographies is normalized by $h_0$ ($=$ 420 m).
	}\label{fig03}
\end{figure*}

A link of the real S-SSW to the quasi-static experiment, in which the time scale of $h$ is unrealistically long, may not be clear. Before conducting such a highly idealized simulation, we perform an experiment with the time scale of $h$ being one week, which is a reasonable time scale for Rossby waves causing S-SSWs \citep{Sjoberg:Birner:2012}. 

We first describe the model configuration common to the experiments with short and rather long time scales of $h$. In Section 2c, the effective topography is given by $h_\mathrm{cmp}$ determined from JRA-55, and the wavenumber-2 component of $h_\mathrm{cmp}$ is the most essential for reproducing the vortex splitting. Moreover, Figure \ref{fig02}d suggests that the phase of the wavenumber-2 $h_\mathrm{cmp}$ does not vary much after $t=-4$ days. Therefore, a new effective topography is defined, whose spatial structure is fixed, but amplitude $a(t)$ is varied with time:
\begin{equation}
	h_\mathrm{QSE} \equiv a(t) \times h_0 \; \cos[2(\lambda - \lambda_0)] \; \exp \left[ -\frac{1}{2} \left( \frac{\varphi-\varphi_0}{\varDelta\varphi} \right)^2 \right],
\end{equation}
where
\begin{align}
	a(t) \equiv \begin{cases}
				0 &\text{$(t \le 0)$}, \\
				a_{\max} \times \frac{t}{\varDelta t} &\text{$(0 < t \le \varDelta t)$}, \\
				a_{\max}                                          &\text{$(\varDelta t < t)$}.
			\end{cases}
\end{align}
Note that the subscript QSE stands for quasi-static experiment. Figure \ref{fig03} compares $h_\mathrm{QSE}$ [$a(t)$ = 1] with the wavenumber-2 $h_\mathrm{cmp}$ (at the onset time). The parameters of $h_0$, $\lambda_0$, $\varphi_0$, and $\varDelta\varphi$ are determined to fit $h_\mathrm{QSE}$ into the wavenumber-2 $h_\mathrm{cmp}$, as in the caption of Fig. \ref{fig03}. Hereafter, the magnitude of $h_\mathrm{QSE}$ is normalized by $h_0$ (i.e., the magnitude is denoted by $a$). The other parameters, $a_{\max}$ and $\varDelta t$, will be given in the following subsections.

The initial PV is given by the axisymmetric component of the climatological, barotropic absolute vorticity $\Delta \psi + f$, which is determined by the following two steps. (i) The climatological (three-dimensional) absolute vorticity is obtained by simply averaging the absolute vorticity over the 56 midwinters (December to February) in JRA-55, except for the periods of SSWs. (ii) Its barotropic component is defined by vertically averaging the obtained absolute vorticity with a weight of density over $\theta = 550$ to $1300$ K (at about 22 to 39 km).

\subsection{Experiment with time scale of one week}

\begin{figure*}[!t]
	\centerline{\includegraphics[width=39pc,angle=0]{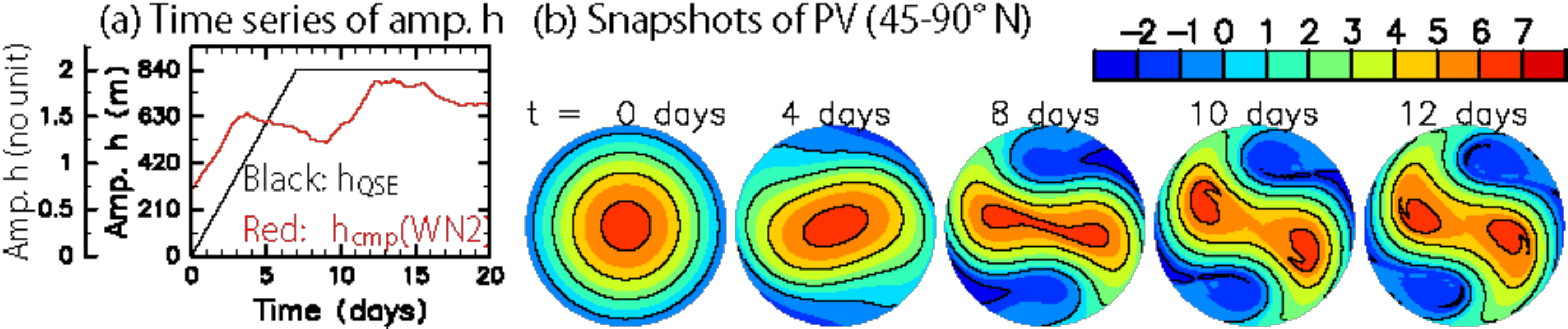}}
	\caption{
	(a) Time series of the topographic amplitudes given by (red) the zonal-wavenumber-2 component of $h_\mathrm{cmp}$ made from JRA-55 and by (black) $h_\mathrm{QSE}$ in (3) and (4), where $\varDelta t = 7$ days and $a_{\max}=2$. The same wavenumber-2 $h_\mathrm{cmp}$ as that in Fig. \ref{fig02}d is used for the red time series. For comparison, the onset time is shifted to $t = 10$ days. (b) Evolution of PV over 45 to 90${}^\circ$ N simulated by the QG model using $h_\mathrm{QSE}$ in (a).
	}\label{fig04}
\end{figure*}

In this subsection, we perform an experiment with the time scale $\varDelta t$ $=$ 7 days, where $a_{\max}$ in (4) is set to $2$. Figure \ref{fig04}a compares the time series of $a(t)$ in (4) with that of the amplitude of the wavenumber-2 $h_\mathrm{cmp}$ developed from JRA-55. The time scale and amplitude of $h_\mathrm{QSE}$ are similar to those of the wavenumber-2 $h_\mathrm{cmp}$.

Figure \ref{fig04}b shows the simulated PV evolution. Compared with Figs. \ref{fig02}c and \ref{fig02}e, a similar vortex splitting is reproduced even by using the simpler effective topography $h_\mathrm{QSE}$ in (3) and (4). This result motivates us to further investigate the transitions in the QG model by performing an experiment with a large $\varDelta t$ (i.e., the amplitude of $h_\mathrm{QSE}$ is slowly increased).

\subsection{Quasi-static experiment with time scale of $O(10^4 \; \mathrm{days})$}

\begin{figure*}[!t]
	\centerline{\includegraphics[width=39pc,angle=0]{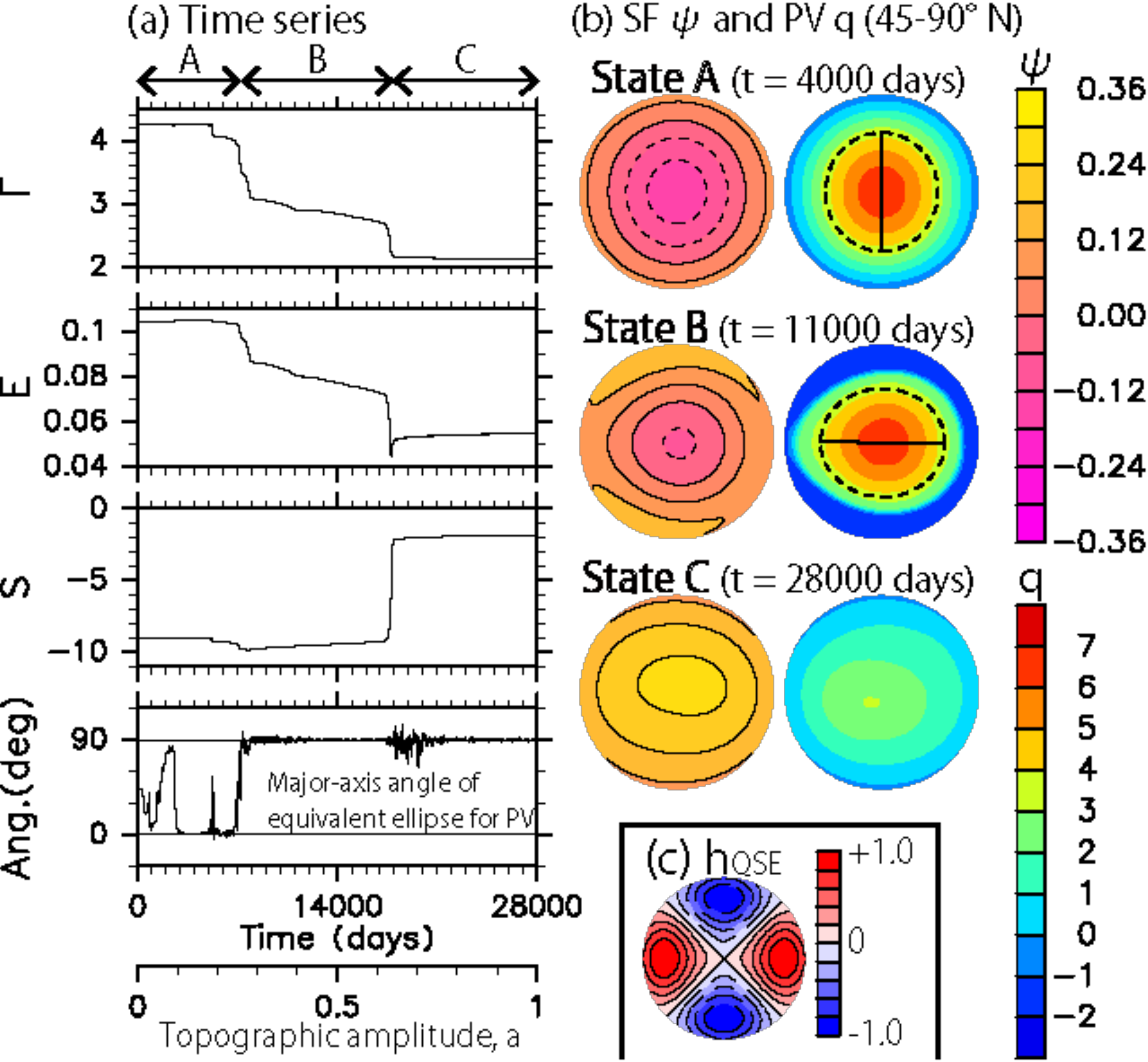}}
	\caption{
	Results of the quasi-static experiment with the QG model using $h_\mathrm{QSE}$ in (3) and (4), where $\varDelta t = 2.8 \times 10^4$ days and $a_{\max}=1$. (a) Time series of the total PV $\Gamma$, the total energy $E$, the negative of potential enstrophy $S$, and the major-axis angle of the equivalent ellipse for PV. The quantities of $\Gamma$, $E$, and $S$ are defined in (5), (6), and (7), respectively. The equivalent ellipse for PV is obtained by following \citet{Matthewman:Esler:2011}, where its major-axis angle is measured from the vertical axis in the clockwise direction. A low-pass filter with a 100-day cutoff period was applied to all time series (the results are insensitive to the cutoff period). (b) Snapshots over $45$ to $90{}^\circ$ N of the three nearly steady states observed in the quasi-static experiment. SF means stream function $\psi$. In the top and middle PV snapshots, the equivalent ellipses are drawn by the dashed curves and their major axes by the solid lines. In (a), the duration of each nearly steady state is roughly denoted by the arrow. (c) Effective topography $h_\mathrm{QSE}$ with $a(t)=1$ over $45$ to $90{}^\circ$ N. The parameters of $h_\mathrm{QSE}$ are the same as those in Fig. \ref{fig03}b, except for $\lambda_0=0^\circ$.
	}\label{fig05}
\end{figure*}

We perform a quasi-static experiment, where $\varDelta t$ and $a_{\max}$ in (4) are set to $2.8 \times 10^4$ days and $1$, respectively. The following results are insensitive to $\varDelta t$. Similar results were obtained when $\varDelta t = 1.4 \times 10^4$ days, which indicates that the variation in the topographic amplitude $a(t)$ is sufficiently slow. Without loss of generality, $\lambda_0$ in (3) is set to zero (see Figs. \ref{fig03}b and \ref{fig05}c).

We examine the evolution of the PV field and the following integrated quantities:
\begin{align}
	\Gamma &= \int \; q \; \mathrm{d}A, \\
	E &= \frac{1}{2} \int \; (\mathbf{\nabla}\psi)^2 \; \mathrm{d}A, \text{ and } \\
	S &= -\frac{1}{2} \int \; q^2 \; \mathrm{d}A,
\end{align}
where $\mathrm{d}A$ is the area element, $\Gamma$ is the total PV, $E$ is the total energy, and $S$ is the negative of potential enstrophy. Significant changes in the PV field are observed over the North Pole, and we focus on the polar cap north of 45${}^\circ$ N. The surface integrals in (5) through (7) are taken over this polar cap.

The PV field is nearly steady, except around the two transitions\footnote{See a supplemental file (spp1.gif) for the animation of the evolution of PV and stream function from the start to the end of the quasi-static experiment.}. Figure \ref{fig05} shows (a) the time series of $\Gamma$, $E$, $S$, and major-axis angle of the equivalent ellipse for PV, (b) the typical snapshots of the three nearly steady states, and (c) the structure of the effective topography $h_\mathrm{QSE}$. The changes in the PV field are correlated with the abrupt changes in $\Gamma$, $E$, and $S$. The first transition occurs at about $t=7000$ days. The state before the transition is named A, in which the polar vortex is vertically elongated (i.e., the major-axis angle $\sim 0^\circ$), as shown in the 4000th-day snapshots. The state after the transition is named B, in which the polar vortex is laterally elongated (i.e., the major-axis angle $\sim 90^\circ$), as shown in the 11 000th-day snapshots. Associated with the transition from A to B, the values of $\Gamma$, $E$, and $S$ rapidly decrease. Note that the temporary increase in the major-axis angle around $t=5000$ days (Fig. \ref{fig05}a) was not observed in some experiments using a different viscosity coefficient or truncation wavenumber, and it is not discussed here.

The major-axis angle of the equivalent ellipse increases before about $t=2000$ days and then reduces to about $0{}^\circ$ (Fig. \ref{fig05}a). This variation is sensitive to slight changes in the initial PV (not shown). This is likely because the aspect ratio of the equivalent ellipse is close to unity (i.e., the ellipse is almost a circle) and there is a high uncertainty in the direction of its major axis, which will lead to the sensitivity of the major-axis angle. During this period, the initial axisymmetric state changes into A, which is not axisymmetric (Fig. \ref{fig05}b). For simplicity, we regard the initial axisymmetric state as A.  When the topographic time scale $\varDelta t$ in (4) is 7 days and the initial PV is given by that of State A (i.e., the 4000th-day PV in Fig. \ref{fig05}b), a similar vortex splitting as that in Fig. \ref{fig04}b is reproduced (not shown). Furthermore, when $\varDelta t$ is $2.8 \times 10^4$ days (quasi-static) and the initial PV includes all zonal-wavenumber components (not axisymmetric), a state similar to A appears and persists for about 3000 days (not shown). These results imply that the emergence and persistence of State A is not highly sensitive to the initial PV and validate that the initial state is regarded as A.

The second transition occurs at about $t =$ 18 000 days, accompanied by the polar-vortex collapse. The PV filaments are peeled off the polar vortex, and eventually the polar vortex breaks down without splitting\footnote{See a supplemental file (spp2.gif) for the animation of the evolution of PV and stream function around the polar-vortex collapse.}. Before this transition, the flow field is at State B. The state after the transition is named C, in which the weak PV patch is laterally elongated, as shown in the 28 000th-day snapshots (Fig. \ref{fig05}b). Associated with the transition from B to C, the values of $\Gamma$ and $E$ rapidly decrease but that of $S$ rapidly increases due to the strong PV mixing (Fig. \ref{fig05}a). Just after the transition, the small PV patch remains, but it is dissipated at about $t =$ 21 500 days.

More importantly, State B (as well as A) is cyclonic; whereas, C is anticyclonic, as seen in the stream functions (Fig. \ref{fig05}b). Note that C is anticyclonic, independent of the small PV patch left after the polar-vortex breakdown. A similar flow field to that of State C was also obtained in the experiment in the previous subsection, where $\varDelta t = 7$ days, sufficiently after the vortex splitting (not shown). These results indicate that the final state C is not sensitive to the topographic time scale $\varDelta t$, but the vortex splitting and the emergence of State B (i.e., the transition path to the final state C) depend on $\varDelta t$.

\citet{Liu:Scott:2015} conducted the parameter sweep experiments where the topographic amplitude was increased linearly and slowly with time, as in our quasi-static experiment, by using spherical one-layer models (QG and shallow-water) with a wavenumber-2 effective topography. In their experiments, two transitions similar to those in our experiment were observed. The difference from our results is that the flow field corresponding to State B oscillates more strongly (they called it the oscillating regime) in their results. This difference is likely attributable to the angular frequency of the effective topography, which is zero in our experiment.

Finally, we examine causes for the variations in $\Gamma$, $E$, and $S$. Their budgets are governed by the following equations:
\begin{align}
	\frac{\mathrm{d}\Gamma}{\mathrm{d}t} &= -\nu \int \; \Delta^{10} q \; \mathrm{d}A - \int \; \mathbf{\nabla \cdot} (\mathbf{v}q) \; \mathrm{d}A, \\
	\frac{\mathrm{d}E}{\mathrm{d}t}            &= \nu \int \; \psi \Delta^{10} q \; \mathrm{d}A + \int \; \mathbf{\nabla \cdot} \left( \mathbf{v}q\psi + \psi \frac{\partial}{\partial t} \mathbf{\nabla}\psi \right) \; \mathrm{d}A \nonumber\\ &+ \int \; \frac{\psi f}{H} \frac{\partial h_\mathrm{QSE}}{\partial t} \; \mathrm{d}A, \text{ and} \\
	\frac{\mathrm{d}S}{\mathrm{d}t} &= \nu \int \; q\Delta^{10} q \; \mathrm{d}A + \int \; \mathbf{\nabla \cdot} \left(\mathbf{v}\frac{q^2}{2}\right) \; \mathrm{d}A,
\end{align}
where all integrations are taken over 45 to 90${}^\circ$ N. Note that if the viscosity coefficient $\nu$ is zero and if all integrations are taken over the whole sphere, $\Gamma$ and $S$ are conserved, but $E$ can be changed by variations in $h_\mathrm{QSE}$ [i.e., the last term in (9)].

\begin{figure*}[!t]
	\centerline{\includegraphics[width=39pc,angle=0]{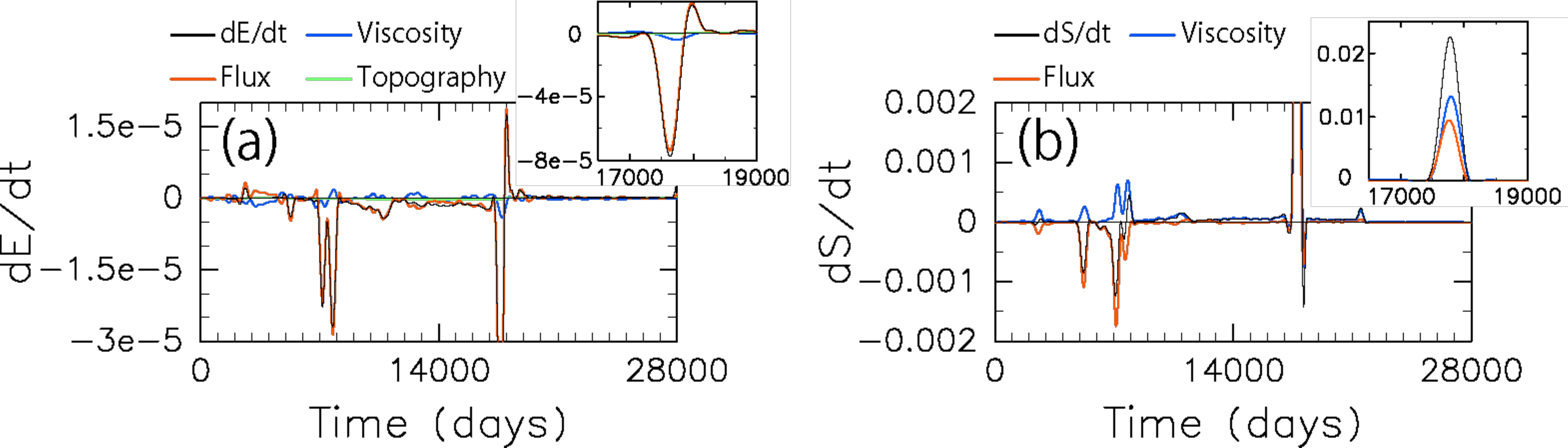}}
	\caption{
	Time series of each term of (a) the energy budget (9) and (b) the enstrophy budget (10) in the quasi-static experiment. The label Viscosity denotes the first terms on the right hand sides of (9) and (10), Flux the second terms of (9) and (10), and Topography the third term of (9). Each inset shows the time series around the polar-vortex collapse. A low-pass filter with a 500-day cutoff period was applied to all time series (the results are insensitive to the cutoff period).
	}\label{fig06}
\end{figure*}

Figure \ref{fig06}a shows the time series of each term in the energy budget (9). Obviously, $E$ is varied at the two transitions ($t \sim$ 7000 and 18 000 days) primarily by the energy fluxes across 45${}^\circ$ N. Similarly, $\Gamma$ is varied primarily by the PV fluxes (not shown). Figure \ref{fig06}b shows the time series of each term in the enstrophy budget (10). The abrupt change in $S$ at the first transition ($t \sim$ 7000 days) is mainly due to the flux term, like $\Gamma$ and $E$, but its change at the second transition ($t \sim$ 18 000 days) is due to the flux term and also to the viscosity term. At the second transition, the small-scale structures develop, the potential enstrophy is transferred into smaller scales, and it is finally dissipated by the numerical viscosity. These results are not highly sensitive to the viscosity coefficient $\nu$. Similar time series to those in Figs. \ref{fig05} and \ref{fig06} were obtained when $\nu$ was ten times as large as the present value.

\section{Theoretical calculations based on equilibrium statistical mechanics}

In the quasi-static experiment, we have observed two abrupt transitions in the large-scale flow-field structure. This is reminiscent of a phase transition in equilibrium statistical mechanics. In this section, we apply statistical-mechanics theory to a barotropic QG system on a disk domain. The main result is that the mean state of the winter stratosphere associated with the cyclonic polar vortex may be regarded as a quasi-stationary state (i.e., a saddle point of the entropy), while the equilibrium state (i.e., the maximum of the entropy) is an anticyclonic flow. In Section 5, the theoretical calculations performed here will be compared with the results of the quasi-static experiment.

\subsection{A variational problem given by equilibrium statistical mechanics}

An inviscid, freely-evolving, two-dimensional flow tends to develop into finer and finer structures, while at a later time it tends to reach a quasi-steady, large-scale coherent structure. Equilibrium statistical mechanics gives a general method for obtaining a large-scale structure realized after strong PV mixing, without describing all the details of the dynamics \citep[e.g.,][]{Salmon:1998, Majda:Wang:2006, Bouchet:Venaille:2012}. Statistical mechanics is a static theory, as it does not give any information on the dynamics, such as the time evolution during a transition. It should also be pointed out that the statistical mechanics applies only to an isolated fluid system without dissipation or forcing. In forced-dissipative cases, the theory may apply only if the inertial time scale (for instance, given by the eddy turnover time) is much smaller than the dissipation and forcing time scales.

The present study employs the following variational problem \citep{Chavanis:Sommeria:1996, Venaille:Bouchet:2009, Venaille:Bouchet:2011b, Naso:etal:2010}, which belongs to a subclass of the more general Miller-Robert-Sommeria theory (Section 1):
\begin{equation}
	\max_{q} \left\{ \; S \; \Big\lvert \; E, \, \Gamma \; \right\},
\end{equation}
where $S$, $E$, and $\Gamma$ are defined in (7), (6), and (5), respectively. Formula (11) means that $S$ is maximized by varying $q$ subject to the two constraints of constant total energy $E$ and total PV $\Gamma$. The solution $q$ of (11) is referred to as the equilibrium state. One of the most important points is that an overwhelming number of possible configurations are associated with the equilibrium state. This means that if one state were picked up at random among all possible states and if a spatial coarse-graining were performed, the equilibrium state would be recovered. Hereafter, the negative of potential enstrophy $S$ is called entropy. The entropy $S$ has a one-to-one correspondence to the mixing entropy, when a probability density function of PV is Gaussian, as in our case considered here \citep{Naso:etal:2010}. A brief explanation of this result is given in Appendix A1.

Any stationary point, i.e., a state with the first variation of $S$ being zero under the two constraints of $E$ and $\Gamma$, satisfies the following linear $q$-$\psi$ relation
\begin{equation}
	q = b {\psi} - c,
\end{equation}
where $b$ and $c$ are Lagrange multipliers, depending implicitly on the two constraints of $E$ and $\Gamma$. A derivation of (12) is given in Appendix A2. This relation means that a streamline is identical to the corresponding PV contour; therefore, any stationary point for (11) is an exact steady solution of the QG system [(1) and (2), where $\nu = 0$]. In addition to the entropy maximum (i.e., the equilibrium state), we focus on a local maximum and a saddle point of $S$, which are referred to as a metastable and a quasi-stationary state, respectively. The equilibrium and metastable states are dynamically and nonlinearly stable against any small-amplitude perturbation, but the quasi-stationary states are not necessarily stable and may be destabilized by some perturbation \citep{Ellis:etal:2002, Venaille:Bouchet:2011b, Naso:etal:2010}. This is the reason for referring to the latter as ``quasi''-stationary states.

\subsection{Equilibrium and quasi-stationary states}

The aim of this subsection is to show that the equilibrium state is anticyclonic; whereas, quasi-stationary states are cyclonic, with a realistic parameter set. Assuming the polar cap over 45 to 90${}^\circ$ N to be a closed domain, we propose now to compute the equilibrium and quasi-stationary states within this domain by solving the variational problem (11). These states are easily expressed in terms of Laplacian eigenmodes \citep{Chavanis:Sommeria:1996, Venaille:Bouchet:2009, Venaille:Bouchet:2011b, Naso:etal:2010}. However, computing the Laplacian eigenmodes over a polar cap (i.e., a part of the sphere) leads to unnecessary technical difficulties because the eigenmodes do not have a simple analytic expression, such as spherical harmonics. For this reason, we consider a simpler geometry, namely a disk, which is obtained by projecting the polar cap north of 45${}^\circ$ N onto the plane with Lambert's azimuthal equal-area projection: $x = \sqrt{2(1-\sin \varphi)} \cos \lambda$ and $y = \sqrt{2(1-\sin \varphi)} \sin \lambda$. The values of $\Gamma$, $E$, and $S$ are invariant under Lambert's projection because they are the surface integrals of scalars, and Lambert's projection preserves the area element. This point facilitates the comparisons of the theoretical calculations with the simulation results (Section 5).

\begin{figure*}[!t]
	\centerline{\includegraphics[width=39pc,angle=0]{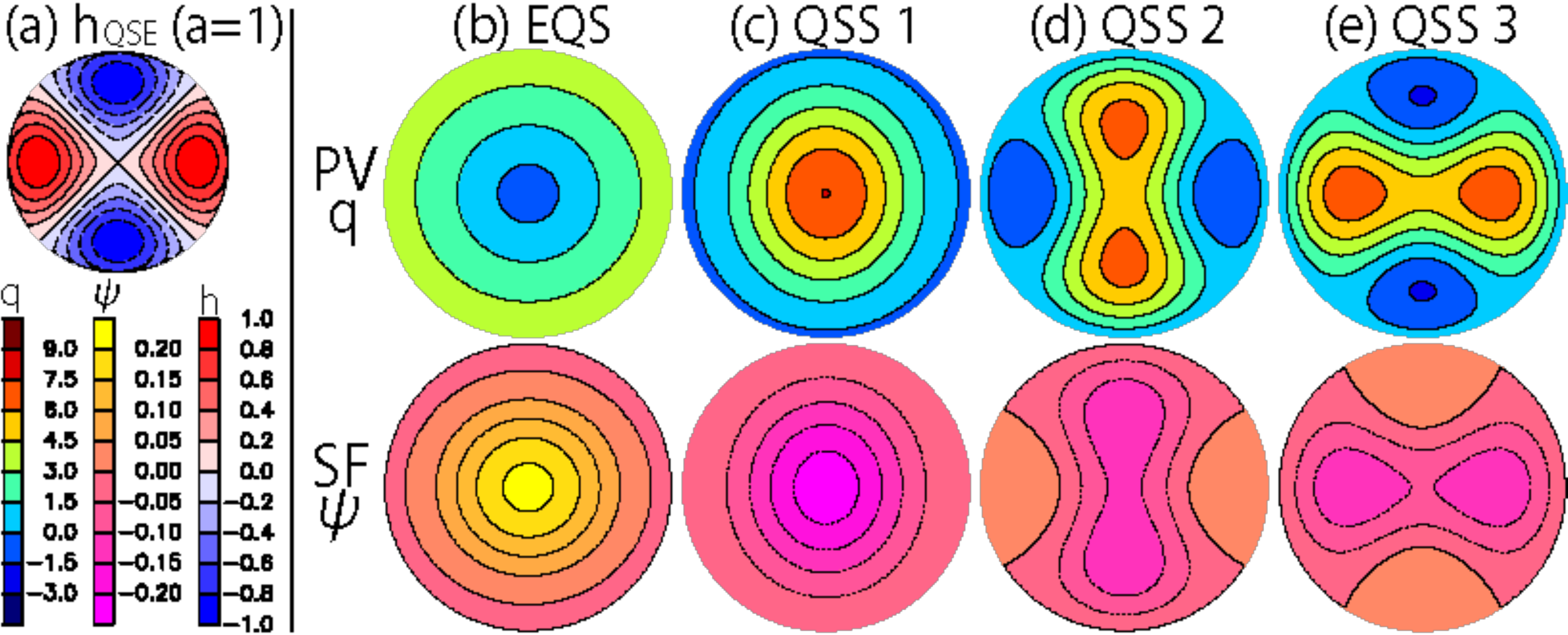}}
	\caption{
	(a) Effective topography $h_\mathrm{QSE}$ over 45 to 90${}^\circ$ N projected onto the plane with Lambert's azimuthal equal-area projection. (b) through (e) Theoretically calculated PV $q$ and stream function (SF) $\psi$ of (b) the equilibrium state (EQS), (c) QSS 1, (d) QSS 2, and (e) QSS 3, where QSS {\it n} stands for the quasi-stationary state having the $n$-th largest $b$ in (12). The parameters are fixed to $(a, \, \Gamma, \, E) = (0.15, \, 4.2, \, 0.10)$.
	}\label{fig07}
\end{figure*}

The Coriolis parameter $f$ and the effective topography $h_\mathrm{QSE}$ are also projected with Lambert's map. The projected $f$ is a monotonic function having a maximum at the origin [$(x, \, y) = (0, \, 0)$], which corresponds to the North Pole. The effect of the earth curvature is partially taken into account, even though we consider a disk on a plane. The spatial structure of $h_\mathrm{QSE}$ is fixed, while its amplitude is controlled by $a$ [see (3)]. The projected $h_\mathrm{QSE}$ with $a=1$ is shown in Fig. \ref{fig07}a. 

A stationary point for the variational problem (11) is determined by the three parameters%
\footnote{Because time does not matter for the variational problem (11), it is always possible to choose a time unit such that $E=1$. Therefore, we are left with only two parameters, the topographic amplitude $a$ and the total PV $\Gamma$. However, the transformation of the time unit makes it a little difficult to compare the theoretical calculations with the simulation results. Therefore, we consider the three parameters of $a$, $\Gamma$, and $E$.}
, the topographic amplitude $a$, the total PV $\Gamma$, and the total energy $E$. The calculation method to obtain the equilibrium and quasi-stationary states is described in Appendix B, and the expressions of the Laplacian eigenmodes are described in Appendix C.

The three parameters are fixed to $(a, \, \Gamma, \, E) = (0.15, \, 4.2, \, 0.10)$ for the moment, which are close to the values taken on the 4000th day in the quasi-static experiment (Fig. \ref{fig05}a). This suggests that these values are in a parameter range relevant to the winter stratosphere. Figure \ref{fig07}b shows the equilibrium PV and stream function. The PV is minimum at the origin (corresponding to the North Pole), but the stream function is maximum there, which means that the equilibrium state is anticyclonic. Similar results are obtained for the other parameters covered by the quasi-static experiment. Therefore, the mean state of the winter stratosphere accompanied by the cyclonic polar vortex may not be considered as an equilibrium state.

\citet{Venaille:Bouchet:2011b} and \citet{Naso:etal:2010} pointed out that a cyclonic state may exist as a quasi-stationary (or metastable) state, even when the equilibrium one is anticyclonic. For the variational problem (11), there are infinitely many quasi-stationary states, where small-scale structures become dominant, as the inclination $b$ in (12) is decreased \citep{Chavanis:Sommeria:1996}. We investigate here a few quasi-stationary states having large $b$. Figures \ref{fig07}c to \ref{fig07}e show the three quasi-stationary states, where QSS {\it n} stands for the quasi-stationary state having the $n$-th largest $b$. The three quasi-stationary states are cyclonic, as seen in their stream functions. The PV patches of QSS 1 and 2 are vertically long, but that of QSS 3 is laterally long. QSS 1 has the largest structure due to the largest $b$, and its PV is maximum at the origin (corresponding to the North Pole). Therefore, QSS 1 appears to be similar to the mean state of the winter stratosphere, where the polar region is covered with the cyclonic polar vortex.

QSS 1 is a special saddle point of the entropy $S$, but QSS 2 and 3 are saddle points without such a property. Expressing the entropy in terms of a quadratic form, we can investigate the structure of the entropy surface in the phase space (see Appendix D for details). Roughly speaking, QSS 1 is virtually a local maximum of $S$, and it is dynamically stable against almost all small-amplitude perturbations. More precisely, in the phase space, the entropy of QSS 1 is increased only along the two directions of the gravest Laplacian eigenmodes with azimuthal wavenumber 1. In other words, QSS 1 may be destabilized only when a perturbation has these wavenumber-1 components. Note that there are two different eigenmodes due to the degeneracy in a disk domain (Appendix C). Moreover, in a general domain without symmetry, such as a rectangular domain, QSS 1 can be a local maximum of the entropy, i.e., metastable (not shown). In this case, QSS 1 is dynamically stable against {\it any} small-amplitude perturbation.

\subsection{Domains of existence of quasi-stationary states}

The results in the previous subsection are at one parameter point [$(a, \, \Gamma, \, E) = (0.15, \, 4.2, \, 0.10)$]. Similar results are obtained over the parameter range of the quasi-static experiment; however, some quasi-stationary states do not exist in some parameter domains. In this subsection, we examine the parameter domains for the existence of QSS 1, 2, and 3.

\begin{figure*}[!t]
	\centerline{\includegraphics[width=39pc,angle=0]{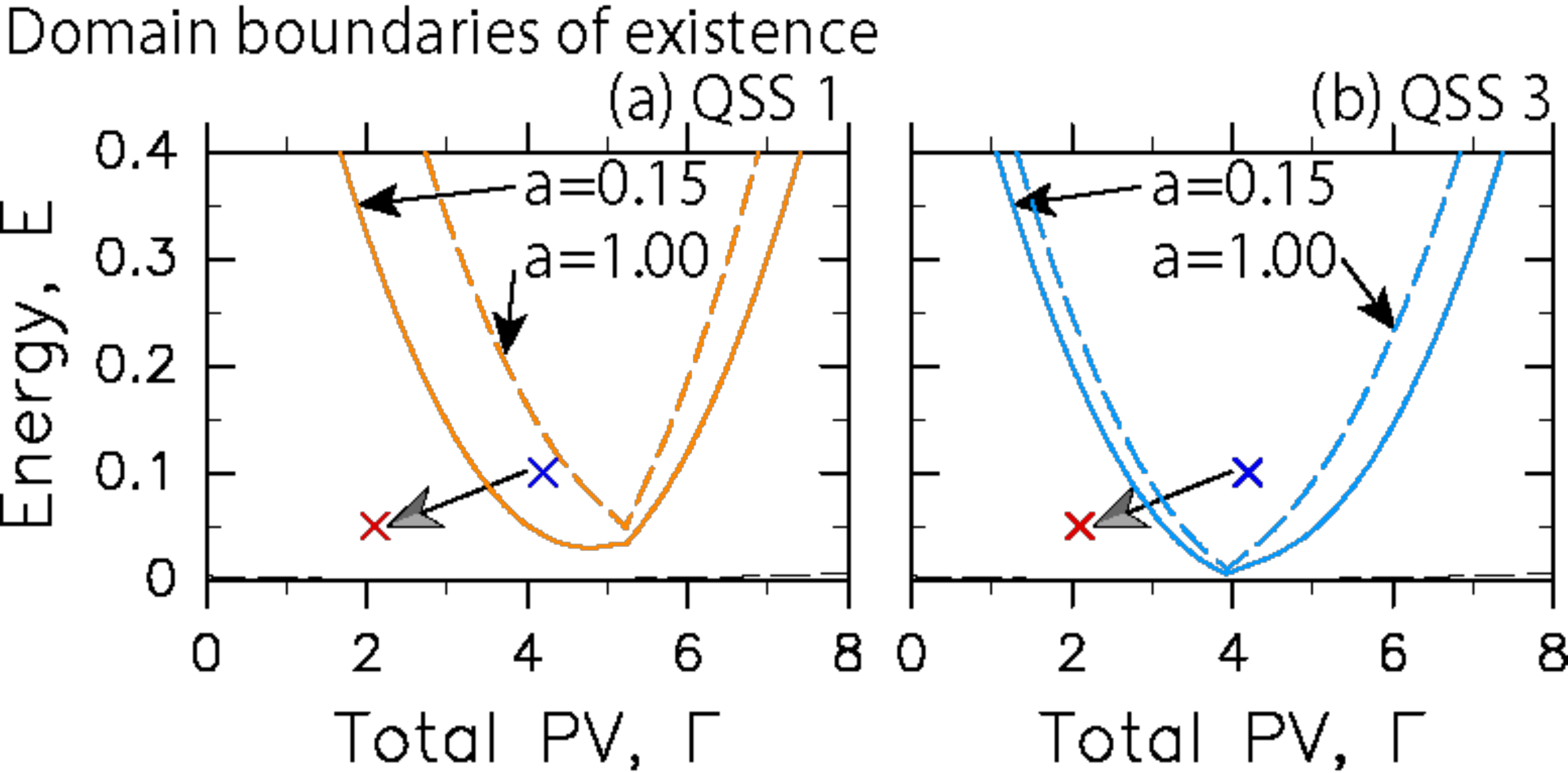}}
	\caption{
	Theoretically calculated domain boundaries of existence of (a) QSS 1 (orange) and (b) QSS 3 (blue) in the $\Gamma$-$E$ space with the topographic amplitude $a=0.15$ (solid) and $a=1.00$ (dashed). The domain boundary for the equilibrium state is also shown by the dashed black curve, where $a=1.00$. Each state exists at higher energies than those on the boundary. In each figure, the blue cross is at $(\Gamma, \, E) = (4.2, \, 0.10)$, and the red cross is at $(\Gamma, \, E) = (2.1, \, 0.05)$.
	}\label{fig08}
\end{figure*}

Figure \ref{fig08} shows the domain boundaries of existence of QSS 1 and 3 in the $\Gamma$-$E$ space. Each state exists at higher energies than those on the colored curves (solid for $a=0.15$ and dashed for $a=1.00$). The method of obtaining these boundaries is described in Appendix B. The domain of existence of QSS 2 is the same as that of QSS 1, because both states annihilate at the same parameters. The domains of existence of QSS 1 and 3 become narrower and shift to regions with higher energies as the topographic amplitude $a$ is increased. The domain boundary for the equilibrium state is also shown by the dashed black curve, which nearly overlaps with the $\Gamma$ axis in each figure. This black curve is almost independent of $a$, and the curve with $a=1.00$ is shown.

In Fig. \ref{fig08}, the blue cross at $(4.2, \, 0.10)$ is close to the values of $(\Gamma, \, E)$ on the 4000th day in the quasi-static experiment (Fig. \ref{fig05}a); whereas, the red cross at $(2.1, \, 0.05)$ is close to those on the 28 000th day. When $\Gamma$ and $E$ vary from the blue to the red cross, the domain boundaries of existence of QSS 1 and 3 are crossed. In this case, if the initial state is QSS 1 or 3, a transition will occur because the parameters enter the domain without both states. It is worth noting that equilibrium statistical mechanics does not give any information on time evolution, such as the relaxation toward the equilibrium state. This also means that the statistical mechanics does not predict the variations of the total PV $\Gamma$ and energy $E$, while these variations are obtained by solving the QG equations and then externally used in the statistical-mechanics theory. A careful analysis of simulation results is necessary to discuss a transition between the two states. We perform this analysis in the next section.

\section{Interpretations of the quasi-static experiment in terms of equilibrium statistical mechanics}

In this section, the results of the quasi-static experiment are compared with the theoretical calculations performed in the previous section. In Section 5a, we first confirm that the statistical-mechanics theory can be applied to the polar cap over 45 to 90${}^\circ$ N. In Section 5b, the PV fields in the quasi-static experiment are qualitatively compared with those of the equilibrium and quasi-stationary states. In Section 5c, quantitative comparisons are made. We demonstrate that a transition occurs when the parameters ($a, \Gamma, E$) enter the domain without an appropriate quasi-stationary state. Finally, in Section 5d, the new interpretation of S-SSWs is proposed.

\subsection{Preliminaries}

The statistical-mechanics theory, which predicts a steady state realized after strong PV mixing, applies only to an isolated system whose inertial time scale is much shorter than the dissipation and forcing time scales. As we stated in Section 3b, the PV field in the quasi-static experiment is not highly sensitive to the topographic time scale $\varDelta t$ (i.e., forcing time scale) or to the viscosity coefficient $\nu$, which controls the dissipation time scale. This implies that the inertial time scale (i.e., the eddy turnover time of the polar vortex) is sufficiently shorter than the dissipation and forcing time scales. In addition, the PV field is nearly steady, except around the two transitions. More precisely, the components of the PV field with time scales of 100 days or less are negligible (not shown). In other words, despite the efflux of PV and energy, the polar cap north of 45${}^\circ$ N can be regarded as a closed domain at a time scale of 100 days or less. Therefore, statistical-mechanics theory, namely the variational problem (11), can be used to interpret changes in the flow-field structure over the North Pole.

\subsection{Qualitative comparisons in terms of PV fields}

In this subsection, we qualitatively compare the PV fields and suggest that the initial state A is interpreted as QSS 1, the intermediate state B as QSS 3, and the final state C as the equilibrium state. As typical snapshots, the 4000th-, 11 000th-, and 28 000th-day PV fields (regarded as A, B, and C, respectively) in Fig. \ref{fig05}b are compared with the theoretically obtained PV fields. 

\begin{figure*}[!t]
	\centerline{\includegraphics[width=39pc,angle=0]{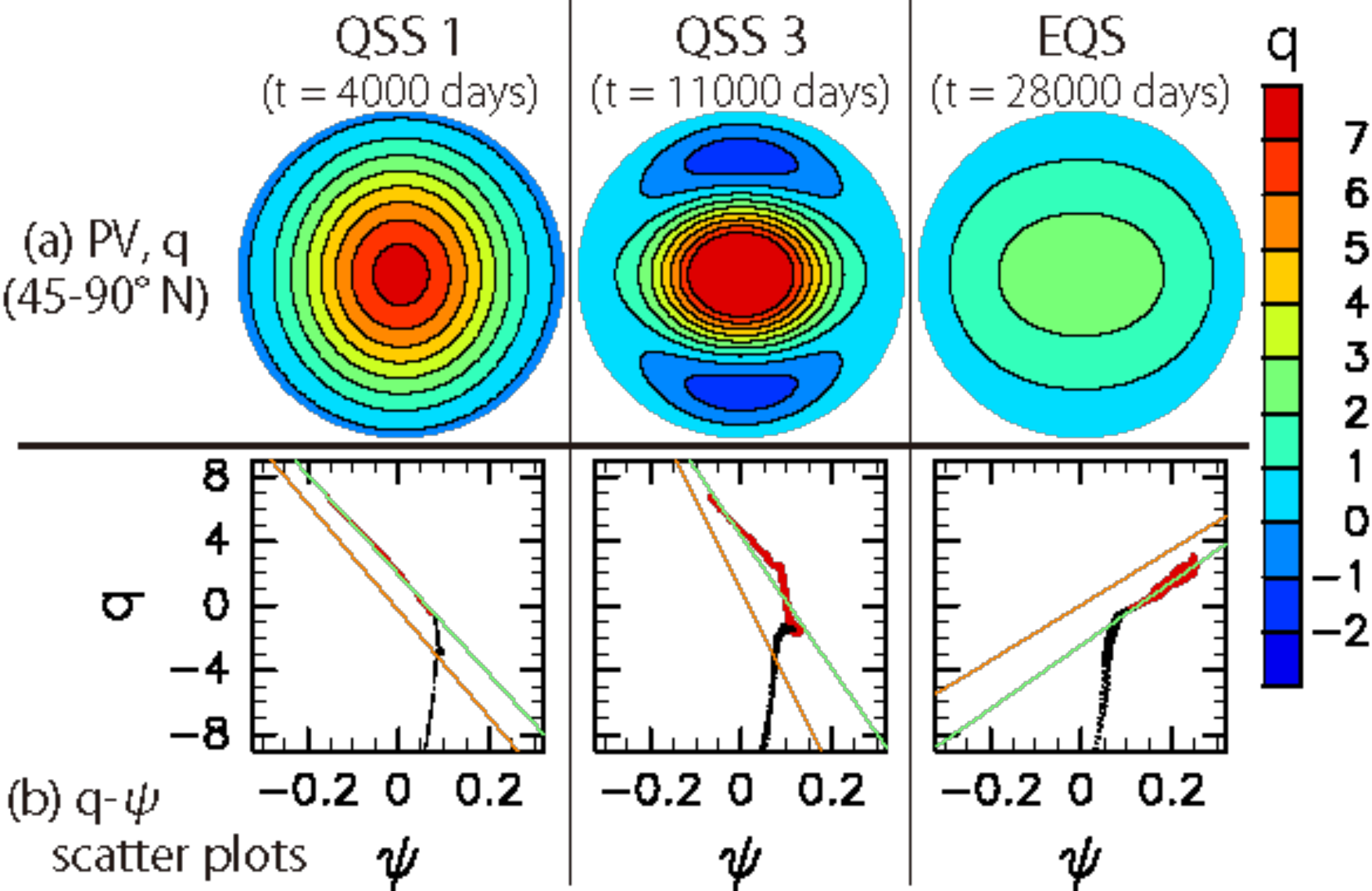}}
	\caption{
	(a) Theoretically calculated PV fields. Each field is obtained by giving the instantaneous parameters of $(a, \, \Gamma, \, E)$ in the quasi-static experiment. For instance, to obtain QSS 1 in the left panel, the parameters of $(a, \, \Gamma, \, E)$ on the 4000th day are used in the theory. Each PV field is projected into the sphere with the inverse of Lambert's map, and then it is projected again onto the plain with an orthographic projection, as for all PV fields in Fig. \ref{fig05}b. (d) $q$-$\psi$ scatter plots in the quasi-static experiment at (left) $t=4000$ days, (middle) 11 000 days, and (right) 28 000 days. The red dots represent the values from the grid points north of 45${}^\circ$ N, and the black ones represent the values from the other points. The green lines are given by the least squares fitting to the red dots. Each orange line represents the $q$-$\psi$ relationship (12) for the above theoretically calculated field.
	}\label{fig09}
\end{figure*}

State A is cyclonic and has a large-scale structure (Fig. \ref{fig05}b), which suggests that A is regarded as QSS 1, because QSS 1 has the largest structure among all quasi-stationary states (Section 4b). Figure \ref{fig09}a shows the PV fields obtained by the theoretical calculations where the parameters of $(a, \, \Gamma, \, E)$ from the quasi-static experiment are used. As expected, the PV magnitude and distribution of QSS 1 are quite similar to those of State A.

State B is cyclonic, and its PV field is slightly elongated along the lateral direction (Fig. \ref{fig05}b), which implies that B can be interpreted as QSS 3, because QSS 3 has the largest structure among all the quasi-stationary states having laterally long PV fields. The PV magnitude of QSS 3 is comparable to that of State B (Figs. \ref{fig05}b and \ref{fig09}a). Although the PV patch of QSS 3 is smaller than that of State B, both PV fields have qualitatively similar shapes.

Only State C is anticyclonic (Fig. \ref{fig05}b), which indicates that C is considered as the equilibrium state, because the equilibrium state is anticyclonic in the parameter range covered by the quasi-static experiment (Section 4b). In fact, the magnitude and distribution of the equilibrium PV field are quite similar to those of State C (Figs. \ref{fig05}b and \ref{fig09}a).

There is also a good agreement on the $q$-$\psi$ relation. Figure \ref{fig09}b shows the $q$-$\psi$ scatter plots obtained from the quasi-static experiment. In each plot, the green line is given by the least squares fitting to the red dots, which represent the values from the grid points north of 45${}^\circ$ N (the black dots show the values from the other points). There are well-defined $q$-$\psi$ relationships over the North Pole, and the linear fitting is valid as the first-order approximation to these relationships. The theory gives the linear $q$-$\psi$ relationship, namely $q = b \psi - c$ in (12), which characterizes the equilibrium and quasi-stationary states. Figure \ref{fig09}b also shows the theoretically calculated linear $q$-$\psi$ relationship (orange lines). The inclinations $b$ given by the theory agree well with those of the quasi-static experiment, but there is a discrepancy in the offsets $c$. This discrepancy is likely because $c$ is dependent on the latitude at the polar-cap boundary. It would be possible to determine a polar cap at each time so that a theoretical value of $c$ is close to a value from the quasi-static experiment. However, such treatment is not necessary for the qualitative comparisons here.

Therefore, we expect that the state changes observed in the quasi-static experiment are understood as
\begin{align}
\text{ QSS 1 (State A)} & \rightarrow \text{ QSS 3 (State B)} \nonumber \\ & \rightarrow \text{ Equilibrium state (State C)}. \nonumber
\end{align}
The polar vortex breaks down without splitting during the transition from B to C. By contrast with the quasi-static experiment, when the topographic amplitude is increased over one week (Section 3a), State B does not appear, and the polar vortex collapses while splitting into the two vortices during the transition from A to C.

\subsection{Quantitative comparisons in terms of transition timings}

The above discussions are qualitative and based on the instantaneous PV snapshots. In this subsection, using the time series of the total PV $\Gamma$ and energy $E$ in the quasi-static experiment, we demonstrate that the timings of the two transitions are consistent with the theoretical predictions. This result supports the expectation of the state changes. The comparisons of transition timings are quantitative and dependent on the size of the polar cap. We properly determine the polar cap for the variational problem (11) and then discuss the transition timings. 

\subsubsection{Determination of polar cap based on surf zone edge}

The theoretical calculations are performed within the disk on the plane, which implies that a discrepancy between the theoretical calculations and the simulation results will be larger as the polar cap becomes larger, due to the difference between the flat and spherical geometries. This suggests that the polar cap should be as small as possible.

The theory describes a flow field having the linear $q$-$\psi$ relationship: $q = b \psi - c$ in (12). If $q$ is nearly constant over some region, only $b=0$ is possible (i.e., $q = c = \mathrm{const}$) because $\psi$ is generally not constant even in such a region. Clearly, $b=0$ is inappropriate for the flow field over the North Pole, as seen in Fig. \ref{fig09}b. The region with nearly constant PV is well-known as the surf zone in the stratosphere \citep[e.g.,][]{McIntyre:Palmer:1983, McIntyre:Palmer:1984}, and a similar region is observed here. Therefore, we define the polar-cap boundary as the north edge of the surf zone. 

\begin{figure*}
	\centerline{\includegraphics[width=39pc,angle=0]{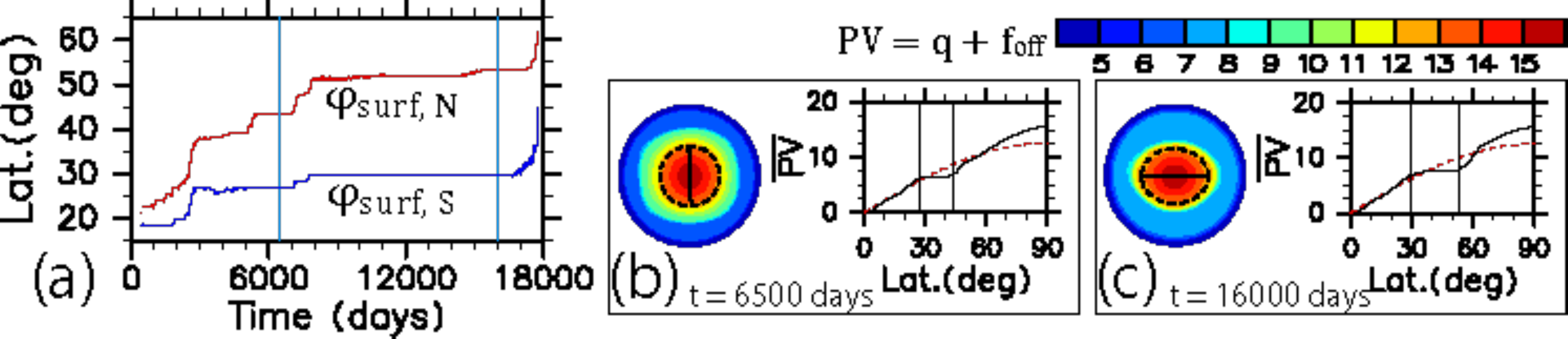}}
	\caption{
	(a) Time series of the north and south edges of the surf zone, $\varphi_\mathrm{surf, N}$ and $\varphi_\mathrm{surf, S}$, respectively, in the quasi-static experiment. The light-blue vertical lines represent $t =$ 6500 and 16 000 days and the label Lat. means latitude (${}^\circ$ N). (b) and (c) PV fields over the Northern Hemisphere and plots of zonal-mean PV versus latitude at (b) $t =$ 6500 and (c) 16 000 days. In the PV snapshots, the equivalent ellipses are drawn by the dashed curves and their major axes by the solid lines. In the zonal-mean PV plots, the red dashed curves represent the Coriolis parameter $f(\varphi) \equiv 2\Omega \sin \varphi$, and the vertical lines represent $\varphi_\mathrm{surf, N}$ and $\varphi_\mathrm{surf, S}$. The PV offset $f_\mathrm{off}$ in (2) is added to $q$: PV $\equiv q+f_\mathrm{off}$.
	}\label{fig10}
\end{figure*}

The north edge of the surf zone is defined in two steps: (i) the latitude with the minimum $\partial \overline{q} / \partial \varphi$ is determined and denoted by $\varphi_{\min}$, where $\overline{q}$ is the zonal-mean PV; and (ii) the north (south) edge of the surf zone $\varphi_\mathrm{surf, N}$ ($\varphi_\mathrm{surf, S}$) is defined as the latitude north (south) of $\varphi_{\min}$ at which $\partial \overline{q} / \partial \varphi$ first exceeds 7.0. The value of 7.0 is arbitrary, but the following results are insensitive to this value. Figure \ref{fig10}a shows the time series of $\varphi_\mathrm{surf, N}$ and $\varphi_\mathrm{surf, S}$, which end at $t = $ 18 000 days due to the polar-vortex breakdown. Figures \ref{fig10}b and \ref{fig10}c show the PV fields and plots of zonal-mean PV, where the vertical lines represent $\varphi_\mathrm{surf, N}$ and $\varphi_\mathrm{surf, S}$. Obviously, this definition of the surf-zone edges captures the region with nearly constant PV. The surf zone becomes wider as the topographic amplitude $a$ is increased. Similar results were reported by \citet{Polvani:etal:1995}, who investigated the surf zone with a spherical shallow-water model including an effective topography.

\subsubsection{Comparisons of transition timings}

\begin{figure*}[!t]
	\centerline{\includegraphics[width=39pc,angle=0]{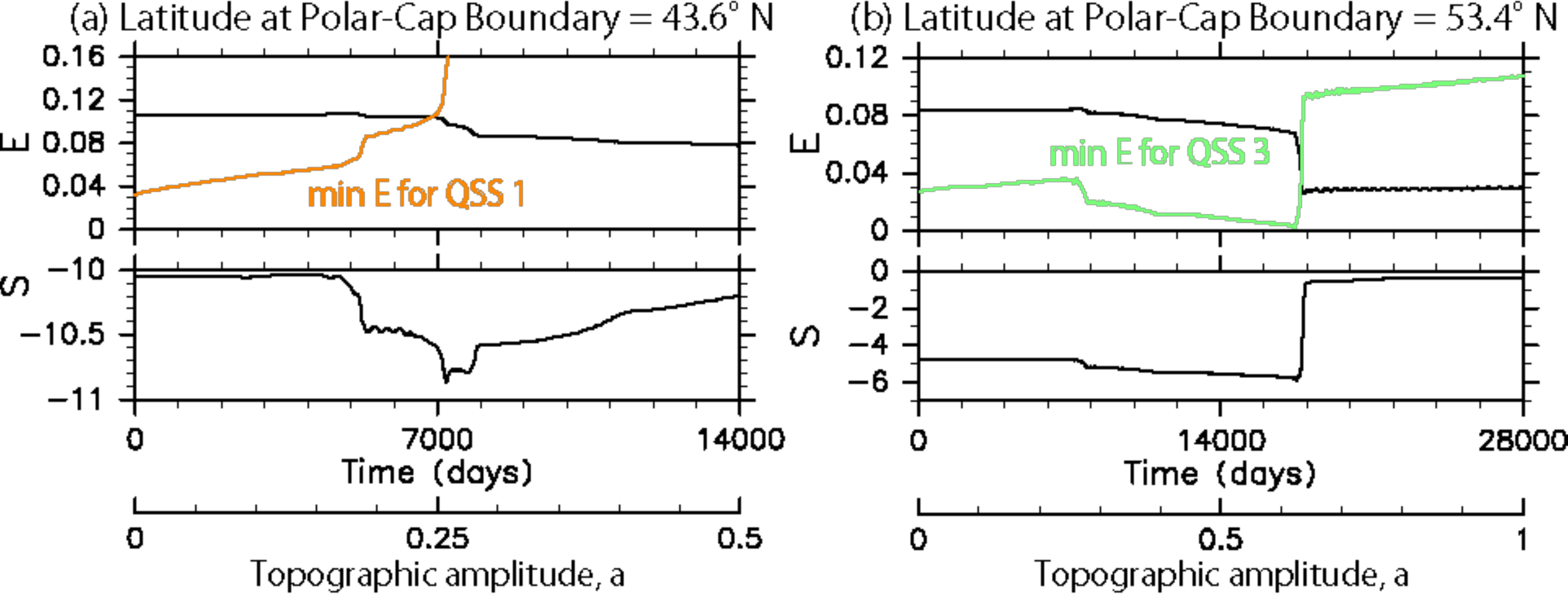}}
	\caption{
	(a) Time series obtained from the quasi-static experiment: (top) total energy $E$ and (bottom) entropy $S$. The orange curve shows the minimum energy for QSS 1, which is theoretically calculated by giving the instantaneous values of the topographic amplitude $a$ and total PV $\Gamma$ in the quasi-static experiment. All calculations are performed within the polar cap north of 43.6${}^\circ$ N. (b) as in (a), but for the minimum energy for QSS 3, where the latitude at the polar-cap boundary is 53.4${}^\circ$ N.
	}\label{fig11}
\end{figure*}

We next discuss the transition timings. The transition from State A to B is first investigated. This transition occurs at about $t = 7000$ days (Fig. \ref{fig05}a), and $\varphi_\mathrm{surf, N}$ is equal to about 43.6${}^\circ$ N just before the transition (Figs. \ref{fig10}a and \ref{fig10}b). Therefore, the theoretical calculations are performed within the polar cap north of 43.6${}^\circ$ N. The minimum energy for the existence of any quasi-stationary state can be theoretically calculated when the topographic amplitude $a$ and the total PV $\Gamma$ are given, as shown in Fig. \ref{fig08}. Since we expect that the transition from A to B is considered as that from QSS 1 to 3, the minimum energy for QSS 1 is calculated at each time by giving the instantaneous values of $a$ and $\Gamma$ in the quasi-static experiment. Figure \ref{fig11}a compares the time series of the minimum energy for QSS 1 (orange) with that of the energy in the quasi-static experiment (black). The bottom panel shows the time series of the entropy $S$ in the quasi-static experiment. In Fig. \ref{fig11}a, the surface integrals of $\Gamma$, $E$, and $S$ in (5) through (7) are taken over 43.6 to 90${}^\circ$ N, and the offset of PV [$f_\mathrm{off}$ in (2)] is given by the Coriolis parameter at 43.6 ${}^\circ$ N. Before reaching the entropy minimum, the energy becomes lower than the minimum energy for QSS 1, which suggests that a transition occurs at about that time. In fact, associated with the decrease in the entropy, the PV field changes from a vertically to a laterally elongated shape (i.e., from State A to B), as shown in the time series of the major-axis angle of the equivalent ellipse (Fig. \ref{fig05}a). These results support that State A before the first transition is regarded as QSS 1.

The entropy minimum is important here because QSS 1 is virtually a local maximum of the entropy (Section 4b). The only way to increase the entropy of QSS 1 is to add a perturbation including the gravest wavenumber-1 modes. In the quasi-static experiment, however, the initial PV is axisymmetric and the effective topography $h_\mathrm{QSE}$ in (3) consists of only the wavenumber-2 component. This means that a wavenumber-1 perturbation is never produced in the QG model, except for a noise from numerical errors. Assuming that such a noise is not critical, we can regard QSS 1 as a local maximum of the entropy. In this case, the entropy will decrease in a transition from QSS 1 to another quasi-stationary state, and the entropy minimum implies that the transition is completed. We confirmed that the results here were hardly changed when a wavenumber-1 perturbation was added to the initial PV in the quasi-static experiment (not shown). Furthermore, the dynamical stability of QSS 1 (State A) was numerically investigated by directly adding a wavenumber-1 perturbation to an intermediate state regarded as QSS 1 and examining its time evolution with the spherical QG model (1) and (2). The small but finite amplitude wavenumber-1 perturbation does not grow with time (not shown). These results imply that wavenumber-1 perturbations are not critical to the emergence and persistence of QSS 1, even though the nonlinear stability of QSS 1 against these perturbations is not theoretically assured in the disk domain (Section 4b).

The second transition from State B to C is examined in a similar way. This transition occurs at about $t =$ 18 000 days (Fig. \ref{fig05}a), and $\varphi_\mathrm{surf, N}$ is equal to about 53.4${}^\circ$ N just before the transition (Figs. \ref{fig10}a and \ref{fig10}c). Since we expect that the transition from B to C is considered as that from QSS 3 to the equilibrium state, the minimum energy for QSS 3 is theoretically calculated within the polar cap north of 53.4${}^\circ$ N and is compared with the energy in the quasi-static experiment. Figure \ref{fig11}b shows the result, as Fig. \ref{fig11}a. The predicted timing of the transition is nearly the same as that in the quasi-static experiment, which is characterized by an abrupt increase in the entropy. The result supports that State B before the second transition is interpreted as QSS 3.

\subsection{A new interpretation of S-SSWs in terms of equilibrium statistical mechanics}

We have demonstrated that the initial state A (cyclonic) is interpreted as QSS 1, the intermediate state B (cyclonic) as QSS 3, and the final state C (anticyclonic) as the equilibrium state. Although the topographic time scale of one week is relevant to S-SSWs \citep{Sjoberg:Birner:2012}, the comparisons between the statistical-mechanics theory and the quasi-static experiment have revealed that the initial state can be considered as QSS 1, which is virtually a local maximum of the entropy (Section 4b). In contrast to the quasi-static experiment, when the topographic time scale is one week (Section 3a), the polar vortex splits and breaks down during the transition from QSS 1 to the equilibrium state.

Therefore, we propose a new interpretation of S-SSWs in terms of equilibrium statistical mechanics. The S-SSW can be qualitatively understood as the transition from the cyclonic quasi-stationary state, i.e., QSS 1 toward the anticyclonic equilibrium state. The vortex splitting is observed during this transition, which is a nonequilibrium state. Without any external forcing, such as radiative cooling, the anticyclonic equilibrium state would be realized at a sufficiently long time after an S-SSW.

A typical phenomenon understood in a similar way is the transition from supercooled water to ice. Supercooled water suddenly changes into ice when some shock is given. The state of ice is the entropy maximum, but that of supercooled water is a local maximum of the entropy. According to our interpretation, supercooled water corresponds to the state before an S-SSW (i.e., QSS 1). Some shock corresponds to the Rossby-wave (i.e., effective-topography) amplification. Ice corresponds to the state at a sufficiently long time after an S-SSW (i.e., the equilibrium state). The only difference is that QSS 1 is virtually a local maximum of the entropy, but supercooled water is exactly a local maximum. The concept of entropy reveals the analogy between the S-SSW and the transition from supercooled water to ice.

\section{Concluding remarks}

The present study has investigated vortex-split sudden stratospheric warmings (S-SSWs) from the viewpoint of equilibrium statistical mechanics. The S-SSW can be qualitatively interpreted as the transition from a cyclonic quasi-stationary state toward the anticyclonic equilibrium state. A quasi-stationary state is defined as a saddle point of the entropy, and an equilibrium state is defined as the entropy maximum. The transient state with the two split vortices is a nonequilibrium state that appears during the transition from the cyclonic quasi-stationary state toward the anticyclonic equilibrium state. This quasi-stationary state is virtually a local maximum of the entropy, and only a few zonal-wavenumber-1 modes can increase its entropy. Without any external forcing, such as radiative cooling, the anticyclonic equilibrium state would be realized at a sufficiently long time after an S-SSW. Our answers to [Q1] and [Q2] in Section 1 are summarized as follows:

\begin{description}
\item[[Ans. to Q1]] The mean state of the stratosphere accompanied by the cyclonic polar vortex is not close to an equilibrium state (anticyclonic), but is close to the dominant quasi-stationary state, i.e., QSS 1 (cyclonic).

\item[[Ans. to Q2]] The S-SSW can be qualitatively interpreted as a transition from a local entropy maximum to a global maximum. 
\end{description}

The details of the results are as follows.

\begin{enumerate}
\item The spherical, barotropic quasi-geostrophic (QG) model has reproduced the evolution of the composite potential vorticity (PV) obtained from the reanalysis dataset (JRA-55). The effective bottom topography in the model has been given by the composite height field of the 550 K potential temperature surface. The zonal-wavenumber-2 component of the effective topography is the most essential ingredient to retain in order to reproduce the vortex splitting.
\item The theory of statistical mechanics, namely the variational problem (11) \citep[i.e., the energy-enstrophy framework, see][]{Chavanis:Sommeria:1996, Venaille:Bouchet:2009, Venaille:Bouchet:2011b, Naso:etal:2010}, has been applied to the polar cap north of 45$^\circ$ N. The equilibrium state is anticyclonic; whereas, the quasi-stationary states are cyclonic in the parameter range relevant to the winter stratosphere. The quasi-stationary state having the largest spatial structure is virtually a local maximum of the entropy.
\item Theoretical calculations have been compared with the results of the quasi-static experiment, in which the topographic amplitude with zonal wavenumber 2 is increased linearly and slowly with time. The theory gives PV fields similar to those in the quasi-static experiment, and the transition timings predicted by the theory are consistent with those in the quasi-static experiment.
\end{enumerate}

There are at least two important future themes. The first is regarding vortex-displacement SSWs (D-SSWs). It will be interesting to examine whether D-SSWs can be interpreted as a transition in terms of equilibrium statistical mechanics. Since a baroclinic structure is essential for D-SSWs \citep{Matthewman:etal:2009, Esler:Matthewman:2011}, a continuously stratified QG model is necessary for investigation. The three-dimensional version of the variational problem (11) has already been applied to geophysical fluid problems \citep[e.g.,][]{Merryfield:1998, Venaille:2012}. It will be a first step to analyze D-SSWs using this theory. Moreover, the polar vortex rapidly becomes weak at the onset of spring, which is called the stratospheric final warming (SFW), and most of these events are vortex-displacement type \citep{Black:McDaniel:2007}. The SFW, i.e., the change from winter to summer, may be regarded as a transition toward the equilibrium state.

The second important theme is related to radiative cooling. Radiative cooling is vital for the reformation of the polar vortex after an S-SSW \citep[e.g.,][]{Rong:Waugh:2004, Scott:Polvani:2006}. If the transition time scale toward the equilibrium state is much longer than the radiative time scale, the cyclonic polar vortex is re-established before the anticyclonic equilibrium state is organized. However, zonal-mean zonal winds often change from westerly to easterly during an S-SSW, which means that the axisymmetric flow temporarily becomes anticyclonic. This implies that the state of the stratosphere temporarily approaches the anticyclonic equilibrium state. The QG model including a radiative relaxation may be able to be examined by using nonequilibrium statistical mechanics, which is a challenging and interesting subject to be addressed in the future to more properly describe S-SSWs.

\acknowledgments
Y. Yasuda (YY) greatly appreciates the advice from Hiroshi Niino, Hisashi Nakamura, and Yohei Onuki on the presentation of the present study. YY is deeply grateful to Kaoru Sato, Masashi Kohma, Soichiro Hirano, and Arata Amemiya for discussing procedures for analyzing JRA-55. YY is also grateful to Keiichi Ishioka and Izumi Saito for giving comments on the modification of the model codes and operation of the model. We would like to thank Editage (www.editage.jp) for English language editing. The GFD-DENNOU library was used to make all figures. The present study was supported by a Grant-in-Aid for Research Fellow (25$\cdot$8466) of the JSPS and by the Leading Graduate School Program for Frontiers of Mathematical Sciences and Physics (FMSP) (Y. Yasuda). The research leading to these results has received funding from the European Research Council under the European Union's seventh Frame-work Programme (FP7/2007-2013 Grant Agreement No. 616811) (F. Bouchet and A. Venaille).


\appendix[A] 
\appendixtitle{Calculations related to variational problem (11)}

We discuss why the negative of potential enstrophy $S$ in (7) corresponds to the mixing entropy and also give a derivation of the linear $q$-$\psi$ relationship (12). The discussions below basically follow the results of \citet{Naso:etal:2010}. One should also see \citet{Bouchet:2008}, which showed that any solution of the variational problem (11) is a solution of the more general Miller-Robert-Sommeria theory (the converse is not necessarily true).

\section{Relation between potential enstrophy and mixing entropy}

We first introduce a probability density function of PV, $\rho(\mathbf{x}, \sigma)$. The value of $\rho \; \mathrm{d}\sigma \, \mathrm{d}\mathbf{x} $ represents probability to observe microscopic PV whose value is between $\sigma$ and $\sigma+\mathrm{d}\sigma$ in the infinitesimal element of $\mathrm{d}\mathbf{x}$ around the position $\mathbf{x}$. A coarse-grained macroscopic PV field $[q]^\mathrm{macro}$ is given by $[q]^\mathrm{macro} \equiv \int \; \sigma\rho \; \mathrm{d}\sigma$. Practically, a PV field observed in a numerical model should be interpreted as $[q]^\mathrm{macro}$ due to the small-scale dissipation. For simplicity, we do not distinguish between $q$ and $[q]^\mathrm{macro}$.

The mixing entropy is defined as
\begin{equation}
	S_\mathrm{mix} \equiv -\int \; \rho(\mathbf{x}, \sigma) \, \log \rho(\mathbf{x}, \sigma) \; \mathrm{d}\sigma \,\mathrm{d}\mathbf{x},
\end{equation}
which is proportional to the natural logarithm of the number $(\equiv N)$ of possible microscopic configurations (i.e., $S_\mathrm{mix} \propto \log N$). \citet{Naso:etal:2010} showed that a Gaussian distribution is necessary to maximize $S_\mathrm{mix}$ under the three constraints of constant total energy $E$, total PV $\Gamma$, and total microscopic potential enstrophy ${\Gamma_2}^\mathrm{micro}$:
\begin{equation} 
	\Gamma_2^{\mathrm{micro}} \equiv \frac{1}{2} \int \;  \sigma^2 \rho(\mathbf{x}, \sigma) \; \mathrm{d}\sigma \, \mathrm{d}\mathbf{x}.
\end{equation}
The definitions of $E$ and $\Gamma$ are the same as (6) and (5), respectively. The Gaussian distribution is given by
\begin{equation}
	\rho(\mathbf{x}, \sigma) = \frac{1}{\sqrt{4 \pi (\Gamma_2^{\mathrm{micro}} + S) }} \exp \left\{ -\frac{\left[ \sigma - q(\mathbf{x}) \right]^2}{4 (\Gamma_2^{\mathrm{micro}} + S) } \right\},
\end{equation}
where $S$ is defined by (7). The distribution $\rho$ in (A3) satisfies the constraint of $\Gamma_2^{\mathrm{micro}} = (1/2) \int \;  \sigma^2 \rho \; \mathrm{d}\sigma \, \mathrm{d}\mathbf{x} \; (= \mathrm{const})$. Substituting (A3) into (A1), we obtain the one-to-one correspondence between $S_\mathrm{mix}$ and $S$:
\begin{equation}
	S_\mathrm{mix} = \frac{1}{2} + \frac{1}{2} \log (4\pi) + \frac{1}{2}  \log \left( \Gamma_2^{\mathrm{micro}} + S \right).
\end{equation}
Since the natural logarithm function is strictly increasing and $\Gamma_2^{\mathrm{micro}}$ is constant, $S$ has to be maximized by varying $q$ in order to maximize $S_\mathrm{mix}$, which leads to the variational problem (11). 

An important point for deriving (A4) is the Gaussian distribution in (A3). This result may be roughly understood as follows. When we calculate the first-order variation of some functional including $S_\mathrm{mix}$, this variation will have a term proportional to $\log \rho$ originating from the variation of $S_\mathrm{mix}$. In addition, when this functional includes $\Gamma_2^{\mathrm{micro}}$, the obtained variation will also have a term proportional to $\sigma^2$. Assuming that the total variation is zero and does not include higher-order terms than $\sigma^2$, we will obtain a Gaussian distribution, such as (A3). The mean value of the Gaussian distribution (A3) is also important, which is equal to the (macroscopic) PV $q(\mathbf{x})$. This mean value comes from a constraint on the (macroscopic) PV. See \citet{Naso:etal:2010} for details.

\section{Derivation of linear $q$-$\psi$ relationship (12)}

According to the Lagrangian multiplier theory \citep[e.g.,][]{Gelfand:Fomin:2000}, the following first-order variation with respect to any $\delta q$ is zero at a stationary point for the variational problem (11):
\begin{equation}
	\delta S - b \delta E - c \delta \Gamma = 0,
\end{equation}
where $\Gamma$, $E$, and $S$ are defined in (5), (6), and (7), respectively, and $b$ and $c$ are Lagrange multipliers. With the aid of integration by parts, $\delta E$ is changed to
\begin{equation}
	\delta E = \int \; \delta\mathbf{\nabla}\psi \mathbf{\cdot} \mathbf{\nabla}\psi \; \mathrm{d}A = - \int \; \psi \delta \Delta\psi \; \mathrm{d}A = - \int \; \psi \delta q \; \mathrm{d}A.
\end{equation}
Therefore, (A5) is transformed to
\begin{equation}
	\int \; \left( -q + b \psi - c \right) \, \delta q \; \mathrm{d}A = 0,
\end{equation}
which yields the linear $q$-$\psi$ relationship (12).

\appendix[B] 
\appendixtitle{Details of calculations to obtain equilibrium and quasi-stationary states}

We describe the method to obtain the equilibrium and quasi-stationary states for the variational problem (11). The discussions below basically follow the results of \citet{Venaille:Bouchet:2011b}, and the same notations are used. We first introduce the complete, orthonormal basis $\{e_i\}_{i \in \mathbb{N}}$ of Laplacian eigenmodes on a simply connected domain such as a disk: $\Delta e_i = -\mu_i e_i$, where an eigenvalue $\mu_i$ is positive. Two subspaces are further introduced. One is composed of the Laplacian eigenmodes having zero mean values ($\langle{e_i}'\rangle=0$), and the other is composed of the eigenmodes having non-zero mean values ($\langle{e_i}''\rangle\ne 0$), where $\langle \; \rangle$ denotes the spatial integral over the domain, and $'$ and $''$ emphasize the difference between the two subspaces. In each subspace, the eigenvalues are in ascending order. The Laplacian eigenmodes and eigenvalues on a disk domain are given in Appendix C.

All quantities are expressed in terms of the coordinates $\{ q_i \}$ [the coefficients of PV $q$ ($\equiv \sum_i q_i e_i$)]:
\begin{align}
	\psi_i &= -\frac{q_i - {h_\mathrm{ tot}}_i}{\mu_i}, \\
	\Gamma &= \sum_i q_i \langle e_i\rangle, \\
	E &= \frac{1}{2} \sum_i \frac{ (q_i - {h_\mathrm{ tot}}_i)^2 }{\mu_i}, \text{ and }\\
	S &= -\frac{1}{2} \sum_i q_i^2,
\end{align}
where $h_\mathrm{tot}$ ($\equiv \sum_i {h_\mathrm{tot}}_i e_i$) is the sum of the Coriolis parameter [$f-f_0$ in (2)] and the effective topography [$f h/H$ in (2)] and $\psi_i$ is a coefficient of the stream function $\psi$ ($\equiv \sum_i \psi_i e_i$). Note that (B1) represents the PV inversion (2) given by $q \equiv \Delta \psi + f - f_0 + f h/H \equiv \Delta \psi + h_\mathrm{tot}$. When both $'$ and $''$ are not attached to variables in a summation, the summation is taken over all indices of both subspaces. The expression of PV at a stationary point for the variational problem (11) is obtained by expanding the linear $q$-$\psi$ relationship (12) by the Laplacian eigenmodes:
\begin{equation}
	q_i = \frac{b {h_\mathrm{ tot}}_i - c \mu_i \langle e_i\rangle}{\mu_i + b}.
\end{equation}
The expressions of $\Gamma$ and $E$ at a stationary point are obtained by substituting (B5) into (B2) and (B3), respectively:
\begin{align}
	\Gamma &= -c F(b) + b \sum_i \frac{{h_\mathrm{ tot}}_i \langle e_i\rangle}{\mu_i + b}, \text{ and}\\
	E &= \left( \sum_i \frac{\mu_i {h_\mathrm{ tot}}_i^2 }{2(\mu_i + b)^2} \right) + c \left( \sum_i \frac{\mu_i {h_\mathrm{ tot}}_i \langle e_i\rangle}{(\mu_i + b)^2} \right) \nonumber \\ &+ c^2 \left( \sum_i \frac{\mu_i \langle e_i\rangle^2}{2(\mu_i + b)^2} \right),
\end{align} 
where
\begin{equation}
	F(b) \equiv \sum_i \frac{\mu_i \langle e_i\rangle^2}{\mu_i + b}.
\end{equation}

Any stationary point given by (B5) with $b > -{\mu_1}'$ and $-\mu^\ast$ is the solution of (11), i.e., the equilibrium state, where ${\mu_1}^\prime$ is the smallest Laplacian eigenvalue for the zero-mean eigenmodes and $-\mu^\ast$ is the largest zero of $F(b)$ [i.e., $F(-\mu^\ast)=0$]. In the parameter range considered in the present study, $b$ of the equilibrium states is always larger than $-{\mu_1}'$ and $-\mu^\ast$. See \citet{Venaille:Bouchet:2011b} for the method to calculate the equilibrium state with $b = -{\mu_1}'$ or $-\mu^\ast$. Note that equilibrium states with $b < -{\mu_1}'$ or $-\mu^\ast$ do not exist.

An equilibrium or a quasi-stationary state is obtained by following four steps: (i) the Lagrange multiplier $c$ is analytically obtained by solving (B6) with a given $\Gamma$; (ii) by substituting the obtained $c$ into (B7), the equation $E=E(b)$ is derived; (iii) the Lagrange multiplier $b$ satisfying $E=E(b)$ is numerically calculated with a given $E$; and (iv) $q_i$ is computed by substituting the obtained $b$ and $c$ into (B5). When $b > -{\mu_1}'$ and $-\mu^\ast$, the obtained $q$ is the equilibrium state. In the other cases, the obtained $q$ is a quasi-stationary state. To calculate a quasi-stationary state, an appropriate $b$ needs to be selected in the third step (iii). For instance, to obtained QSS 1, we select $b$ at the intersection of the first branch of the energy curve $y = E(b)$ with the line of $y = E$. If such an appropriate $b$ does not exist, the corresponding quasi-stationary state does not exist. In preparing Fig. \ref{fig08}, we numerically checked whether an appropriate $b$ existed at each parameter point of $(a, \Gamma, E)$. In other words, a domain with existence of a quasi-stationary state is a parameter set over which the corresponding branch of the energy curve $y = E(b)$ has an intersection with $y=E$.

\appendix[C]
\appendixtitle{Laplacian eigenvalues and eigenmodes in a disk domain}

We give the Laplacian eigenmodes and eigenvalues in a disk domain. A position in the domain is specified by a radius $r$ and an azimuthal angle\footnote{The azimuthal angle is identical to longitude $\lambda$ when a disk domain is obtained with Lambert's map.} $\lambda$, where the maximum of $r$ is designated as $r_{\max}$. The Laplacian eigenmodes are given by Bessel functions of the first kind $J_n$ and trigonometric functions when the Dirichlet boundary condition is imposed (i.e., $e_i = 0$ at $r=r_{\max}$):
\begin{align}
	\{ {e_i}' \}_{i \in \mathbb{N}} &= \biggl\{ {A}_{n,m} \; J_n \left( \frac{\alpha_{n, m}}{r_{\max}} r \right) \sin(n\lambda), \nonumber \\ &\quad\quad {A}_{n,m} \; J_n \left( \frac{\alpha_{n, m}}{r_{\max}} r \right) \cos(n\lambda) \biggr\}_{n, m \in \mathbb{N}}, \\
	\{ {e_i}'' \}_{i \in \mathbb{N}} &= \left\{ {A}_{0,i} \; J_0 \left( \frac{\alpha_{0, i}}{r_{\max}} r \right) \right\}_{i \in \mathbb{N}}, \text{ and } \\
	 \langle {e_i}'' \rangle &= \frac{2 \sqrt{\pi} r_{\max}}{\alpha_{0, i}} \mathrm{sgn}\left( J_1\left( \alpha_{0, i}\right) \right),
\end{align}
where ${A}_{n,m}$ is a normalization constant ($\langle (e_i)^2 \rangle = 1$) and $\alpha_{n, m}$ is the $m$-th zero of $J_n$. The corresponding eigenvalues $\mu$ are given by $(\alpha_{n, m}/r_{\max})^2$.

\appendix[D]
\appendixtitle{Quadratic form expressing entropy around a stationary point}

According to the Lagrangian multiplier theory \citep[e.g.,][]{Gelfand:Fomin:2000}, the necessary and sufficient condition that a stationary point for the variational problem (11) is a local maximum of the entropy is given by the following second-order variation, in which a perturbation $\delta q$ satisfies the two first-order constraints:
\begin{align}
	0 < - \delta^2 {S} + b \delta^2 {E} &= \frac{1}{2} \sum_i \left( 1 + \frac{b}{\mu_i} \right) (\delta q_i)^2 \nonumber\\ &\text{ s.t. } \; \delta {\Gamma}=0 \; \text{ and } \; \delta {E} = 0.
\end{align}
The first-order constraints are expressed in terms of $\{ \delta q_i \}$:
\begin{align}
	0 &= \delta {\Gamma} = \sum_i \langle e_i\rangle \delta q_i = \sum_i \langle{e_i}^{\prime\prime}\rangle \delta {q_i}^{\prime\prime}, \text{ and} \\
	0 &= \delta {E} = \sum_i \psi_i \delta {q_i},
\end{align}
where $\delta q_i$ is a coefficient of $\delta q$ by Laplacian eigenmodes (see also Appendix B), and $\psi_i$ is given by substituting (B5) into (B1). 

The necessary and sufficient condition (D1) is not quite easy to handle because of the two first-order constraints. We solve these two linear constraints [(D2) and (D3)] for ${\delta q_1}^{\prime\prime}$ and ${\delta q_2}^{\prime\prime}$: %
\begin{align}	
	{\delta q_1}'' &= \nonumber \\ & \sum_{i \ge 3} \left[ -\frac{\langle{e_i}''\rangle}{\langle{e_1}''\rangle} +  \frac{\langle{e_2}''\rangle}{\langle{e_1}''\rangle} \left( \frac{{\psi_i}''-{\psi_1}''\langle{e_i}''\rangle/\langle{e_1}''\rangle}{{\psi_2}'' - {\psi_1}''\langle{e_2}''\rangle/\langle{e_1}''\rangle}  \right) \right] {\delta q_i}'' \nonumber \\
	&+ \sum_{i \ge 1} \frac{\langle{e_2}''\rangle}{\langle{e_1}''\rangle} \left( \frac{{\psi_i}'}{{\psi_2}'' - {\psi_1}''\langle{e_2}''\rangle/\langle{e_1}''\rangle}  \right){\delta q_i}' \nonumber \\
	&\equiv \sum_{i} {B_i} {\delta q_i}, \text{ and } \\
	{\delta q_2}'' &= - \sum_{i \ge 3} \left( \frac{{\psi_i}''-{\psi_1}''\langle{e_i}''\rangle/\langle{e_1}''\rangle}{{\psi_2}'' - {\psi_1}''\langle{e_2}''\rangle/\langle{e_1}''\rangle}  \right) {\delta q_i}'' \nonumber \\
	&- \sum_{i \ge 1} \left( \frac{{\psi_i}'}{{\psi_2}'' - {\psi_1}''\langle{e_2}''\rangle/\langle{e_1}''\rangle}  \right){\delta q_i}' \nonumber \\
	&\equiv \sum_{i} {C_i} {\delta q_i}.
\end{align}
Substituting ${\delta q_1}''$ and ${\delta q_2}''$ into (D1), we obtain the quadratic form expressing the second-order variation (D1), in which the two first-order constraints are incorporated:
\begin{align}
        &2 \left( - \delta^2 {S} + b \delta^2 {E} \right) = \sum_{\mathrm{other} \; i'} {{\delta q_i}'}^2 \left( 1 + \frac{b}{{\mu_i}'} \right) + \nonumber \\
	&\sum_{i,j} \underbrace{ \left[ \delta_{ij} \left( 1 + \frac{b}{{\mu_i}} \right) + B_i B_j \left( 1 + \frac{b}{{\mu_1}''} \right) + C_i C_j \left( 1 + \frac{b}{{\mu_2}''} \right) \right] }_{Q_{ij}} \nonumber\\ &\times {\delta q_i} {\delta q_j},
\end{align}
where $\delta_{ij}$ is a Kronecker delta and other $i'$ means all indices of the zero-mean eigenmodes whose ${\psi_i}'$ are zero (i.e., whose $B_i$ and $C_i$ are zero). The quadratic form (D6) is decomposed into the first sum consisting of the diagonal matrix and the second sum consisting of the symmetric matrix $Q$. In the present study, the second sum is given by the components having wavenumber 0 or 2, and the other components contribute to the first sum. The quadratic form (D6) expresses the entropy surface around a stationary point in the phase space. When (D6) is positive definite, the condition (D1) is satisfied, and the stationary point is a local maximum of the entropy (i.e., dynamically and nonlinearly stable). 

The definiteness of the quadratic form (D6) is examined in the following three steps: (i) the definiteness of $Q$ is checked by numerically computing the eigenvalues of $Q$; (ii) the value of $b$ is compared with the eigenvalues ${\mu_i}^{\prime}$, and the first sum in (D6) is positive for any $\delta q$, if and only if $-{\mu_1}^\prime < b$, where ${\mu_1}$ is the Laplacian eigenvalue of the gravest wavenumber-1 modes [$n=1$ and $m=1$ in (C1)]; and (iii) if $Q$ is positive definite and if $-{\mu_1}^\prime < b$, the quadratic form (D6) is positive definite, and the stationary point is a local maximum of the entropy. Clearly, any stationary point with $b < -{\mu_1}^\prime$ does not satisfy this condition; therefore, it is a saddle point of the entropy. \citet{Naso:etal:2010} showed the same result through a different method. They found a specific first-order perturbation $\delta q$ satisfying the first-order constraints. Substituting it into $- \delta^2 {S} + b \delta^2 {E}$, they showed that the condition (D1) is not satisfied for a stationary point with $b < -{\mu_1}^\prime$.

The uniqueness of QSS 1 comes from the positive definiteness of the symmetric matrix $Q$. This means that QSS 1 is a local maximum of the entropy if $-{\mu_1}^\prime < b$. In other words, the structure of the entropy surface around QSS 1 is determined only by the inequality of $b$ and $-{\mu_1}^\prime$. Furthermore, the value of ${\mu_1}^\prime$ is determined only by the shape of a domain. 

In a disk domain, $-\mu^\ast < -{\mu_1}^\prime$ holds. The range of $b$ for QSS 1 is analytically obtained as $-{\mu_2}^\prime < b < -\mu^\ast$, where ${\mu_2}^\prime$ is the Laplacian eigenvalue of the gravest wavenumber-2 modes [$n=2$ and $m=1$ in (C1)]. Therefore, any $b$ of QSS 1 is smaller than $-{\mu_1}^\prime$, and QSS 1 is always a saddle point of the entropy. In the quadratic form (D6), the gravest wavenumber-1 modes of a perturbation, ${\delta q_1}^\prime$, are included only in the first sum. If a perturbation $\delta q$ does not have these components, the quadratic form (D6) is positive. By contrast, in a rectangular domain, QSS 1 satisfies $-{\mu_1}^\prime < b$, depending on the parameters of $a$, $\Gamma$, and $E$. Therefore, QSS 1 can be a local maximum of the entropy, i.e., metastable (not shown).

\bibliographystyle{ametsoc2014}
\bibliography{references}

\end{document}